\documentclass[twocolumn]{aastex631}

\usepackage[T1]{fontenc}
\usepackage[utf8]{inputenc} 
\usepackage[english]{babel}
\usepackage{amsmath}
\usepackage{booktabs}
\usepackage{graphicx}
\usepackage{multirow}

\newcommand{\rmswhite}{\ensuremath{\sigma_{\tau,\mathrm{wh}}}}

\newcommand{\mbf}[1]{\mbox{\boldmath $#1$}}

\newcommand{\pauli}[1]{\ensuremath{\mbf{\sigma}_{#1}}}
\newcommand{\trace}{{\rm tr}}
\newcommand{\real}{{\rm Re}}

\begin{document}

\title[Instrumental Polarization Calibration]{Reducing instrumental errors in Parkes Pulsar Timing Array data}

\author[0009-0004-7606-9343]{Axl F. Rogers}
\affiliation{Auckland University of Technology,  Private Bag 92006, Auckland 1142, New Zealand}

\author[0000-0003-2519-7375]{Willem van Straten}
\affiliation{Manly Astrophysics, 15/41-42 East Esplanade, Manly, NSW 2095, Australia}

\author[0000-0003-0186-5551]{Sergei Gulyaev}
\affiliation{Auckland University of Technology,  Private Bag 92006, Auckland 1142, New Zealand}

\author[0000-0002-4140-5616]{Aditya Parthasarathy}
\affiliation{Max-Planck-Institut für Radioastronomie, Auf dem Hügel 69, 53121 Bonn, Germany}

\author[0000-0003-1502-100X]{George Hobbs}
\affiliation{CSIRO Astronomy \& Space Science, Australia Telescope National Facility, P.O. Box 76, Epping, NSW, 1710, Australia}

\author[0000-0001-7016-9934]{Zu-Cheng Chen}
\affiliation{Department of Physics and Synergetic Innovation Center for Quantum Effects and Applications, Hunan Normal University, Changsha, Hunan 410081, China}
\affiliation{Institute of Interdisciplinary Studies, Hunan Normal University, Changsha, Hunan 410081, China}

\author[0000-0002-0475-7479]{Yi Feng}
\affiliation{Research Center for Astronomical Computing, Zhejiang Laboratory, Hangzhou 311100, China}

\author[0000-0003-3189-5807]{Boris Goncharov}
\affiliation{Gran Sasso Science Institute (GSSI), I-67100 L'Aquila, Italy}
\affiliation{INFN, Laboratori Nazionali del Gran Sasso, I-67100 Assergi, Italy}

\author[0009-0001-5071-0962]{Agastya Kapur}
\affiliation{CSIRO Astronomy \& Space Science, Australia Telescope National Facility, P.O. Box 76, Epping, NSW, 1710, Australia}

\author[0000-0002-2187-4087]{Xiaojin Liu}
\affiliation{Faculty of Arts and Sciences, Beijing Normal University, Zhuhai 519087, People's Republic of China}

\author[0000-0002-2035-4688]{Daniel Reardon}
\affiliation{Centre for Astrophysics and Supercomputing, Swinburne University of Technology, P.O. Box 218, Hawthorn, Victoria 3122, Australia}
\affiliation{OzGrav: The Australian Research Council Centre of Excellence for Gravitational Wave Discovery, Hawthorn VIC 3122, Australia}

\author[0000-0002-1942-7296]{Christopher J. Russell}
\affiliation{CSIRO Scientific Computing, Australian Technology Park, Locked Bag 9013, Alexandria, NSW 1435, Australia}

\author[0000-0002-9583-2947]{Andrew Zic}
\affiliation{Department of Physics and Astronomy and MQ Research Centre in Astronomy, Astrophysics and Astrophotonics, Macquarie University, NSW 2109, Australia}

\begin{abstract}
This paper demonstrates the impact of state-of-the-art instrumental calibration techniques on the precision of arrival times obtained from 9.6 years of observations of millisecond pulsars using the Murriyang 64-m CSIRO Parkes Radio Telescope. Our study focuses on 21-cm observations of 25 high-priority pulsars that are regularly observed as part of the Parkes Pulsar Timing Array (PPTA) project, including those predicted to be the most susceptible to calibration errors. We employ Measurement Equation Template Matching (METM) for instrumental calibration and Matrix Template Matching (MTM) for arrival time estimation, resulting in significantly improved timing residuals with up to a sixfold reduction in white noise compared to arrival times estimated using Scalar Template Matching and conventional calibration based on the Ideal Feed Assumption. The median relative reduction in white noise is 33\%, 
{and the maximum absolute reduction is 4.5~$\mu$s.}
For PSR~J0437$-$4715, METM and MTM reduce the best-fit power-law amplitude (2.7 $\sigma$) and spectral index (1.7 $\sigma$) 
of the red noise in the arrival time residuals, which can can be tentatively interpreted as mitigation of $1/f$ noise due to otherwise unmodeled steps in polarimetric response. 
These findings demonstrate the potential to directly enhance the sensitivity of 
pulsar timing array experiments through more accurate methods of instrumental calibration and arrival time estimation.
\end{abstract}


\section{INTRODUCTION}\label{sec:INTRODUCTION}

Pulsar Timing Arrays (PTAs) are invaluable tools for detecting spatially correlated signal fluctuations at low frequencies, spanning from nHz to $\mu$Hz \citep{Sazhin_1978, Detweiler_1979, Hellings_Downs_1983}. Within this frequency band, the dominant anticipated signal is the stochastic gravitational-wave background (GWB) generated by a cosmic population of inspiralling supermassive black hole binaries (SMBHBs) \citep{Sesana_2004, Burke-Spolaor_2019}. Additional speculative sources of gravitational waves (GWs) in the nHz range encompass cosmic strings \citep{Siemens_2007, Blanco-Pillado_2018}, phase transitions \citep{Caprini_2010, Kobakhidze_2017, Xue_2021}, and a primordial GWB originating from quantum fluctuations of the gravitational field during the early universe, amplified by inflation \citep{Grishchuk_1975, Lasky_2016}.

The GWB is expected to manifest as a common red noise process, characterized by a similar spectral signature, in all pulsars within the array \citep{Phinney_2001}. Detection of spatially correlated signals in the Times-of-Arrival (ToAs) of multiple pulsars \citep{Rajagopal_Romani_1995}, with the quadrupolar signature (HD correlation) initially proposed by \cite{Hellings_Downs_1983}, provide compelling evidence of a stochastic GWB detection \citep{Agazie_2023c}. Over the past decade, PTA {collaborations} have steadily improved the sensitivity of their data sets for GW searches, progressively reducing the upper limits on the stochastic GWB amplitude \citep{Haasteren_2011, Demorest_2013, Shannon_2013, Arzoumanian_2014, Lentati_2015, Shannon_2015, Arzoumanian_2016, Verbiest_2016, Arzoumanian_2018, Perera_2018, Alam_2021a, Alam_2021b}, and working towards identifying individual GW sources \citep{Yardley_2010, Zhu_2014, Babak_2016, Aggarwal_2019}. 

Beyond GWs, various astrophysical processes can introduce red noise that is unique to each pulsar \citep{Coles_2011, Haasteren_Levin_2013, Lentati_2014}. These include intrinsic spin noise \citep{Shannon_Cordes_2010, Melatos_Link_2014, Lam_2017}, magnetospheric torque variations \citep{Lyne_2010}, variable dispersion \citep{Keith_2013, Jones_2017} and multi-path propagation effects in the inter-stellar medium (ISM) \citep{Hemberger_2008, Cordes_Shannon_2010, Dolch_2021}, and the presence of undetected objects in orbit around pulsars \citep{Shannon_2013b}.

The International Pulsar Timing Array (IPTA) collaboration, composed of various PTA {projects}, is dedicated to the pursuit of a common goal: the detection of nHz-frequency GWs \citep{Hobbs_2010, Perera_2019, Verbiest_2016}. Currently, the IPTA encompasses four PTA members: the European Pulsar Timing Array \citep[EPTA;][]{Kramer_Champion_2013, Desvignes_2016}, the Indian Pulsar Timing Array \citep[InPTA;][]{Joshi_2018, Susobhanan_2020}, the North American Nanohertz Observatory for Gravitational-Waves \citep[NANOGrav;][]{McLaughlin_2013, Arzoumanian_2018, Cordes_2019, Ransom_2019}, and the Parkes Pulsar Timing Array \citep[PPTA;][]{Manchester_2013, Reardon_2016, Kerr_2020}. Furthermore, efforts are underway to establish a PTA project in China (CPTA), leveraging the Five-hundred-meter Aperture Spherical Telescope \citep[FAST;][]{Lee_2016, Hobbs_2019}. Other scientific initiatives, such as the MeerKAT PTA in South Africa \citep{Bailes_2016}, and the Canadian Hydrogen Intensity Mapping Experiment (CHIME) pulsar collaboration in Canada \citep{Ng_2018}, are poised to contribute to the IPTA's collaborative endeavours. An essential addition to this list is the $\gamma$-ray pulsar timing array (Fermi PTA), which offers an independent probe of the GWB and stands as the sole means of confirming radio PTA results \citep{Ajello_2022}.

In 2020 and 2021, significant advancements were made in our understanding of PTAs. The NANOGrav 12.5-year data set \citep[NG12.5;][]{Arzoumanian_2020}, EPTA second data release \citep[EPTA DR2;][]{Chen_2021}, and PPTA second data release \citep[PPTA DR2;][]{Goncharov_2021} all revealed a common uncorrelated red noise (CURN) process within their data sets, marking a crucial discovery. However, these studies failed to conclusively confirm or refute the presence of hypothetical HD correlations.

This revelation gained further support through an analysis of the second data release from the International Pulsar Timing Array \citep[IPTA DR2;][]{Perera_2019}, which consolidated historical data from EPTA, NANOGrav, and PPTA, and confirmed the existence of a CURN process \citep{Antoniadis_2022}. 
However, independent analysis of PPTA DR2 by \cite{Goncharov_2021} shed light on the potential misinterpretation of noise without a statistically identical spectrum between pulsars as a common red process.

PTAs demonstrate sensitivity not only to the quadrupolar correlation of GWs \citep{Taylor_2016, Burke-Spolaor_2019} but also to various other correlated signals. These can include monopolar correlation due to terrestrial time standards errors \citep{Hobbs_2012, Hobbs_2020}, and dipolar correlation stemming from errors in the solar system ephemeris model \citep{Champion_2010, Caballero_2018}. Incorrect modeling of these sources may introduce red noise into timing residuals, potentially compromising GW detection sensitivity \citep{Tiburzi_2016}. While NG12.5 ruled out monopolar and dipolar spatially correlated signals \citep{Arzoumanian_2020}, the analysis of timing residuals in NG12.5, EPTA DR2 and PPTA DR2 do not yield statistical support for quadrupolar spatial correlation \citep{Arzoumanian_2020, Chen_2021, Goncharov_2021}.

The most recent GWB search papers by the EPTA  \citep{Antoniadis_2023}, NANOGrav \citep{Agazie_2023}, and PPTA \citep{Reardon_2023} present analyses of their latest data releases: EPTA second data release \citep[EPTA DR2new+;][]{Antoniadis_2023b}, NANOGrav 15-year data set \citep[NG15;][]{Agazie_2023c}, and PPTA third data release \citep[PPTA DR3;][]{Zic_2023}. These papers unveil evidence for an HD-correlated GWB with varying levels of significance, with estimated probabilities of false alarm rates at 3$\sigma$, 3-4$\sigma$ and 2$\sigma$, respectively. A comparison of these studies showcases consistent measurements of nHz GWB parameters, even with diverse data modeling approaches, demonstrating agreement within 1$\sigma$. The coherence of pulsar noise parameters in the majority of analysed pulsars and the standardisation of noise models reconcile modeling disparities, refining constraints on GWB amplitude and HD correlations. This advancement provides a robust foundation for IPTA's Data Release 3 by extending data sets to encompass additional pulsars \citep{Agazie_2023b}. Concurrently, the Chinese Pulsar Timing Array (CPTA) also reported similar findings on the HD-correlated GWB \citep{Xu_2023}, aligning with the broader consensus from other PTA {projects}.

The sensitivity of PTA experiments is founded upon the accuracy and precision with which arrival times can be estimated. 
Therefore, it remains important to study and quantify the extent to which PTA sensitivity may be limited by unmodeled instrumental artefacts and calibration errors. Instrumental polarisation can distort pulse profiles, leading to correlated errors in the ToAs of each pulsar that can mimic signals associated with a stochastic GWB \citep{Straten_2013,Lentati_2016}. 
Consequently, polarimetric calibration is crucial for minimising systematic timing errors \citep{Straten_2006,gcv+23}, which are most readily observed as dramatic variations of arrival time residuals as a function of parallactic angle \citep[e.g.\ Figure 1 of][]{Straten_2013}. Various methods, including enhanced arrival time estimation \citep{Hotan_2005, Straten_2006}, instrumental calibration \citep{Jenet_Anderson_1998, Straten_2004}, and radio frequency interference (RFI) mitigation \citep{Lazarus_2016, Lazarus_2020, Reardon_2021}, have been developed to quantify and mitigate sources of systematic error, thereby enhancing the precision and accuracy of ToA estimates.

For most pulsars, modest instrumental distortion can induce systematic timing errors of the order of 100 ns \citep{Straten_2013}, significantly hindering efforts to detect the stochastic GWB \citep{Jenet_2005}. In this study, we investigate instrumental distortion of arrival times estimated for the 25 high-priority MSPs regularly observed for the PPTA project, which includes pulsars with great potential for improvement by addressing calibration errors \citep{Straten_2013}.

The paper's structure is as follows: we describe the data set in \S\ref{sec:OBSERVATIONS}, detail our methods in \S\ref{sec:METHODS}, present our results in \S\ref{sec:RESULTS}, and conclude with a discussion of our findings and prospects for future research in \S\ref{sec:DISCUSSION}.

\section{OBSERVATIONS}\label{sec:OBSERVATIONS}

The observations analysed in this study were carried out using the Murriyang 64-m radio telescope at Parkes.
Our analysis focuses on a subset of the data from PPTA DR2, which provides precise arrival times for 26 millisecond pulsars. (PSR~J1732{$-$}5049 was excluded from this research as it was recently removed from the PPTA’s high-priority list.) DR2 spans a 14-year period, with an observational cadence of approximately three weeks, and includes observations from various radio frequency bands (10-cm, 20-cm, 40/50-cm) and backend instruments (CASPSR, CPSR2, PDFBs, WBCORR) \citep{Kerr_2020}. Notably, DR2 yielded the lowest root-mean-square (RMS) timing residuals for each pulsar up to the point of our analysis, which precedes PPTA DR3. 

The selected subset of PPTA DR2 comprises observations conducted between MJDs 55409 and 58933 (from 1 August 2010 to 25 March 2020) at 1400 MHz (corresponding to the 20-cm band observations) employing the H-OH and 21-cm Multibeam receivers and CASPSR backend.
Although CASPSR was also used to observe in 
the 10-cm ($\sim3100$ MHz) and 40/50-cm ($\sim700$ MHz) bands, our analysis focuses solely on the 20-cm band observations due to the limited availability of 10-cm data and the significant RFI contamination in the 40/50-cm band \citep{Parthasarathy_2019}.
Please refer to Table~\ref{tab:pulsar_data} for more details about the observations of each pulsar.

\begin{table*}
    \centering
    \begin{tabular}{c c c c c c}
        \toprule
        Pulsar & $P$ & O$_\mathrm{ToA}$ & N$_\mathrm{ToA}$ & Span & MJD Range \\
        (JNAME) & (ms) & & & (yr) & (start -- finish)\\
        \midrule
        PSR~J0437$-$4715  &  5.76 & 882 & 700 & 8.210 & 55409 -- 58933 \\
        PSR~J0613$-$0200  &  3.06 & 238 & 221 & 8.773 & 55427 -- 58752 \\
        PSR~J0711$-$6830  &  5.49 & 349 & 333 & 9.565 & 55427 -- 58932 \\
        PSR~J1017$-$7156  &  2.34 & 353 & 333 & 8.764 & 55472 -- 58749 \\
        PSR~J1022+1001    & 16.45 & 241 & 222 & 8.172 & 55426 -- 58410 \\
        PSR~J1024$-$0719  &  5.16 & 138 & 127 & 9.098 & 55426 -- 58749 \\
        PSR~J1045$-$4509  &  7.47 & 187 & 177 & 9.051 & 55444 -- 58749 \\
        PSR~J1125$-$6014  &  2.63 & 150 & 141 & 8.865 & 55694 -- 58932 \\
        PSR~J1446$-$4701  &  2.19 & 170 & 145 & 8.913 & 55677 -- 58932 \\
        PSR~J1545$-$4550  &  3.58 & 133 & 123 & 6.621 & 56513 -- 58931 \\
        PSR~J1600$-$3053  &  3.60 & 189 & 174 & 9.551 & 55444 -- 58932 \\
        PSR~J1603$-$7202  & 14.84 & 222 & 212 & 9.095 & 55427 -- 58749\\
        PSR~J1643$-$1224  &  4.63 & 167 & 155 & 9.057 & 55445 -- 58753 \\
        PSR~J1713+0747    &  4.57 & 256 & 246 & 9.059 & 55444 -- 58753 \\
        PSR~J1730$-$2304  &  8.12 & 186 & 165 & 9.598 & 55427 -- 58932 \\
        PSR~J1744$-$1134  &  4.07 & 253 & 237 & 9.057 & 55445 -- 58753 \\
        PSR~J1824$-$2452A &  3.05 &  87 &  80 & 8.923 & 55493 -- 58752 \\
        PSR~J1832$-$0836  &  2.72 &  78 &  72 & 6.804 & 56447 -- 58932 \\
        PSR~J1857+0943    &  5.36 & 138 & 127 & 9.515 & 55457 -- 58932 \\
        PSR~J1909$-$3744  &  2.95 & 374 & 344 & 8.926 & 55444 -- 58752 \\
        PSR~J1939+2134    &  1.56 & 153 & 142 & 7.823 & 55472 -- 58329 \\
        PSR~J2124$-$3358  &  4.93 & 227 & 221 & 9.474 & 55472 -- 58933 \\
        PSR~J2129$-$5721  &  3.73 & 255 & 238 & 9.516 & 55457 -- 58933 \\
        PSR~J2145$-$0750  & 16.05 & 224 & 211 & 9.474 & 55471 -- 58932 \\
        PSR~J2241$-$5236  &  2.19 & 399 & 329 & 9.644 & 55410 -- 58933 \\
        \bottomrule
    \end{tabular}
    \caption{Observational characteristics of 25 high-priority PPTA pulsars, including J2000.0 coordinates (JNAME), spin period ($P$) in milliseconds, observed ToAs (O$_\mathrm{ToA}$), remaining ToAs after outlier rejection (N$_\mathrm{ToA}$), observation time span (in years), and corresponding modified Julian date (MJD) range. Note that for PSR~J1022$+$1001, ToAs from August 15th to 30th each year were excluded in consideration of solar conjunction.}
    \label{tab:pulsar_data}
    
\end{table*}
\section{METHODS}\label{sec:METHODS}

The following sections describe the two methods of polarimetric calibration and the two methods of arrival time estimation that are compared in this work.

\subsection{Polarimetric Calibration}\label{sec:calibration}

To establish a baseline for comparing the impact of calibration techniques, data are calibrated using an estimate of the polarimetric response of the observing system based on an approximation known as the \emph{Ideal Feed Assumption} \citep[IFA;][]{Manchester_2013}.
The IFA calibration model includes the assumptions that the receptors are perfectly orthogonally polarised and that the reference source (e.g., a pulsed noise diode coupled to the receptors) is 100\% linearly polarised, illuminating both receptors equally and in-phase \citep{Caleb_2019}. The IFA is an incomplete description of the instrumental response, and, for some systems, calibration based on the IFA results in significant systematic variations of the total intensity profile and arrival time distortions.  Therefore, for comparison in this study, the data are also calibrated using \emph{Measurement Equation Template Matching} \citep[METM;][]{Straten_2013}. METM uses a single well-calibrated observation of a pulsar with a high signal-to-noise ratio as a polarized reference source, one or more uncalibrated observations of the same pulsar, and (optionally), observations of an amplitude-modulated reference source. 

For this study, PSR~J0437$-$4715 is used as the polarized reference source, and its template profile is derived from multiple MEM solutions. MEM uses uncalibrated observations of a pulsar observed at multiple parallactic angles and an amplitude-modulated reference source. We employ the polarimetric calibration pipeline ({\textsc{PSRPL}})\footnote{http://psrchive.sourceforge.net/manuals/psrpl/} to perform both MEM and METM on multiple data sets. Following RFI excision using \textsc{MeerGuard}\footnote{https://github.com/danielreardon/MeerGuard}, the Meertime extension of \textsc{CoastGuard}\footnote{https://github.com/plazar/coast\_guard} \citep{Lazarus_2016, Lazarus_2020}, we generate five-minute sub-integrations of PSR~J0437$-$4715 and two-minute sub-integrations of noise diode observations. Separate calibrator models are produced for each data set. 

During the MEM stage, eq.~(19) of \cite{Britton_2000} is used to parameterize the {unknown response of the non-ideal} receiver, and the degenerate model parameters described in appendix B of \cite{Straten_2004} are set to zero. 
{The degenerate model parameters include the difference 
in the receptor ellipticities, $\delta_{\chi}$, which mixes 
Stokes~I and Stokes~V, and the rotation of the receiver about 
the line of sight, $\sigma_{\theta}$, which mixes Stokes~Q and 
Stokes~U.  By setting these parameters to zero, the receptors 
are assumed to have equal (and opposite) ellipticities, and 
the rotation of the receiver about the line of sight is assumed 
to be zero.}
{The unknown polarization of the non-ideal noise diode signal is parameterized by the three components of the normalized Stokes polarization vector, $\hat{\mbf C}=(\hat{C}_1,\hat{C}_2,\hat{C}_3)=(C_1,C_2,C_3)/C_0$, where $C_0$ is the total intensity of the noise diode and, for linearly polarized receptors, $C_1$, $C_2$, and $C_3$ correspond to Stokes~Q, U, and V of the noise diode.}

%
%
%
From a total of 66 best-fit MEM solutions, the 52 most robust solutions with a weighted mean reduced $\chi^2$ (averaged over all frequency channels) less than 1.05 and greater than 0.85 are selected. The selected solutions are then ranked based on the product of several range-normalized attributes. {Given the minimum and maximum values, $x_\mathrm{min}$ and $x_\mathrm{max}$, of some attribute $x$, 
and the value of that attribute derived from the $i^\mathrm{th}$ solution $x_i$, 
the (dimensionless) range-normalized attribute,
\begin{equation}
\hat{x}_i = \frac{x_i - x_\mathrm{min}}{ x_\mathrm{max} - x_\mathrm{min} }
\end{equation}
lies on the interval $[0,1]$.  This normalization gives equal weight to each of the attributes included in the rank metric, which is a function of} the integration length of the session, the signal-to-noise ratio (S/N) of the integrated average profile, the weighted mean reduced $\chi^2$, the median uncertainties of the estimated values of $\delta_\theta$ (the difference in receptor orientations) and $\sigma_\chi$ (the receptor ellipticities), and the fraction of the band that was lost to RFI.
%
%
Based on these attributes, the MEM solution derived from the observing session recorded on 2014 April 15 is ranked as the best, and the calibrated total profile observed on this day is selected as the reference profile.  
%
%
%

\begin{figure}
    \centering
    \includegraphics[angle=-90,width=0.46\textwidth]{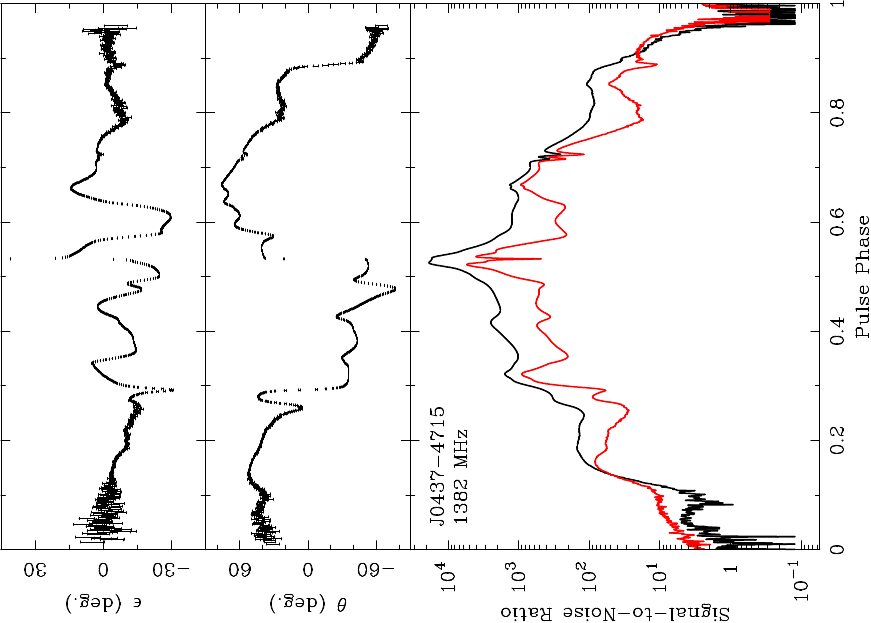}
    \caption{Average polarization of PSR~J0437$-$4715, plotted as a function of pulse phase using polar coordinates: orientation, $\theta$, ellipticity, $\epsilon$, and polarized intensity plotted in red below the total intensity. Flux densities are normalized by the standard deviation of the off-pulse total intensity phase bins. Calibrated using MEM and used as the template for METM, this profile is integrated from 160 hours of observations that span 9.6 years and 400~MHz of bandwidth centred at 1382~MHz.}
    \label{fig:template}
\end{figure}

The 51 other MEM-calibrated total profiles are matched to this reference profile using Matrix Template Matching \citep[MTM;][]{Straten_2006}, then the reference profile and matched totals are integrated to form a template profile with an integration length of 160 hours, shown in Figure~\ref{fig:template}.
%
%
Between pulse phase $\sim 0.9$ and 0.13, the polarized flux appears to be greater than the total intensity, which is not physically possible.  This artefact of imperfect baseline removal may indicate that radiation from this pulsar is received at all times; therefore, there is no off-pulse region of pulse phase, and each of the four Stokes parameters is offset by an arbitrary amount. This baseline artefact does not affect arrival time estimates, which are calculated using only the non-zero spin harmonics of the Fourier transform of each pulse profile.
{Although PSR~J0437$-$4715 is the brightest pulsar with the highest precision in the PPTA, it is also exceptionally susceptible to calibration errors, primarily owing to the transition between orthogonally polarised modes in the middle of its main pulse.}

This template profile is used as the polarized reference source in the following stage of METM analysis, during which eq.~(19) of \cite{Britton_2000} is used to parameterize the receiver and all model parameters are varied.
%
%
A total of 296 robust METM solutions are selected by placing a lower limit on the fraction of the band that was lost to RFI and by applying Tukey's fence thresholds (see \S\ref{sec:timing}) to various attributes, including the median (over all frequency channels) reduced $\chi^2$ and the median uncertainties of the estimated values of $\delta_\theta$, $\sigma_\chi$, $\hat{C}_1$ and $\hat{C}_2$.

%
%

\begin{figure}
%
%
%
%
\centering
    \includegraphics[width=0.48\textwidth]{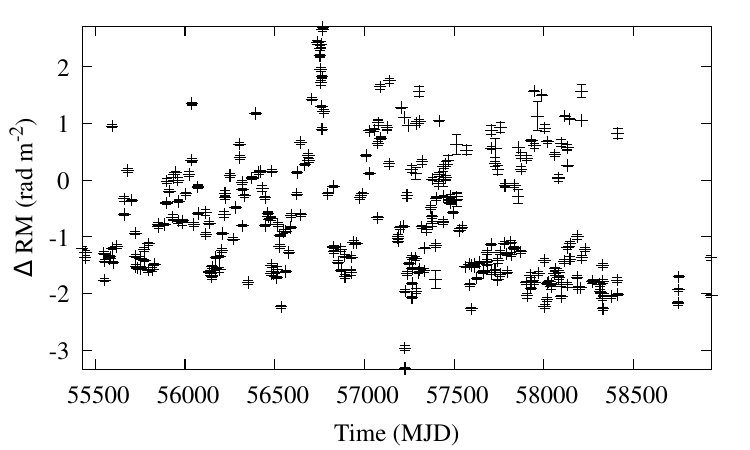}
    \caption{Temporal variation of average daily ionospheric contribution to Faraday rotation, measured with respect to the ionosphere on 2014 April 15 (MJD 56762). Most of the error bars, which denote the 1-$\sigma$ uncertainty of the RM estimate, are too small to be distinguished from the data point.}
    \label{fig:rm}
\end{figure}

In this analysis, variation of ionospheric Faraday rotation is not included in the METM model; therefore, for each METM solution, the variation of the best-fit estimate of $\sigma_\theta$ with radio frequency is used to derive an estimate of the average change in ionospheric Faraday rotation on that day, relative to 2014 April 15.  After excluding outliers, the ionospheric Faraday rotation measure differences ($\Delta$RM shown in Figure~\ref{fig:rm}) vary between $-3.3$ and $+2.7$ rad\,m$^{-2}$ over the 9.6 years spanned by robust METM solutions.  The predominantly day-to-night variation of ionospheric total electron content is observed as annual variations in the derived ionospheric $\Delta$RM estimates owing to the annual drift between solar time and the sidereal times at which PSR~J0437$-$4715 is observed. The observed peak in ionospheric $\Delta$RM around MJD~56750 (2014 March/April) is near the peak in solar magnetic activity cycle 24.
These estimates of ionospheric RM are used to correct the estimates of $\sigma_\theta$ in each METM solution.

\begin{figure}
\centering
    \includegraphics[width=0.52\textwidth]{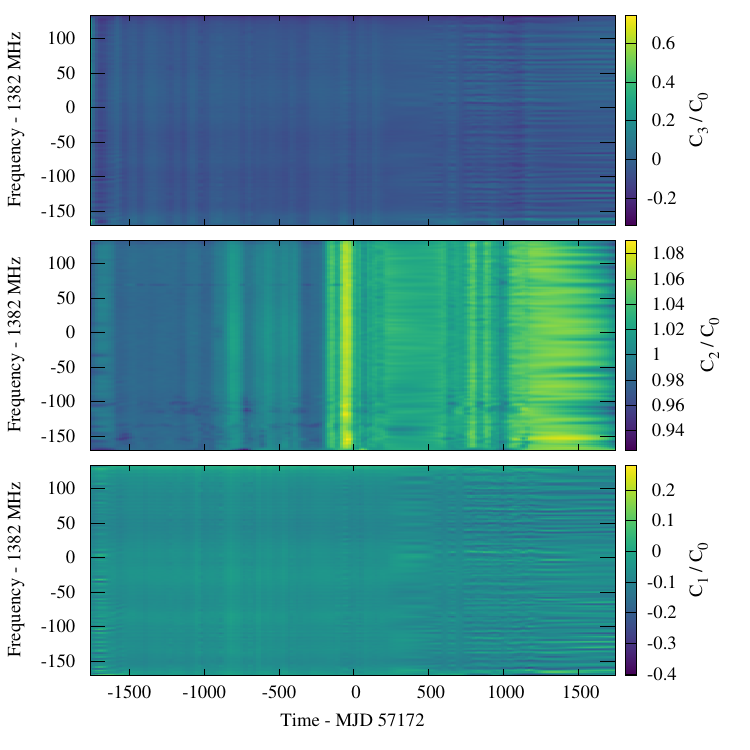}
    \caption{Two-dimensional smoothing splines fit to the best estimates of the {normalized polarization vector of the artificial noise source; each component is} plotted as a function of time and frequency.}
    \label{fig:cal_smint}
\end{figure}

To reduce the noise in the estimated model parameters and interpolate across gaps in both time and radio frequency, seven of the METM model parameters ($\sigma_\theta$, $\sigma_\chi$, $\delta_\theta$, $\delta_\chi$, $\hat{C}_1$, $\hat{C}_2$, and $\hat{C}_3$) are smoothed using two-dimensional penalized splines \citep{10.1214/ss/1038425655,SPLINTER}.
%
%
The optimal smoothing factor is determined using an {iterative search algorithm.
For each trial smoothing factor, the average goodness-of-fit is evaluated over four iterations 
of Monte Carlo cross-validation, also known as repeated random sub-sampling validation.}
On each of the four iterations, the smoothing spline is fit to a randomly selected half 
of the parameter estimates, and the other half of the estimates are used to validate the 
goodness-of-fit of the spline.  The smoothing splines fit to the best estimates of $\hat{\mbf C}$ are shown in Figure~\ref{fig:cal_smint}. We speculate that the apparent over-polarization of the noise diode, which starts around MJD 57000, is of unknown instrumental origin.  In future work, it might prove useful to model instrumental impurity using a depolarizing Mueller matrix \citep{lc96} and include this model in the calibration solution.

Given an observation of the artificial noise source, the components of $\hat{\mbf C}$ predicted by the smoothing splines are used to derive estimates of the absolute gain $G$, differential gain $\gamma$, and differential phase $\phi$ of the instrument as described in \cite{Ord_2004}\footnote{Only $\hat{\mbf C}$ is required to determine $G$, $\gamma$, and $\phi$ because the METM model is configured to include the reference source in the signal chain after the front-end component described by $\sigma_{\theta}$, $\sigma_{\chi}$, $\delta_{\theta}$, and $\delta_{\chi}$.}. 
These are combined with the values of $\sigma_{\theta}$, $\sigma_{\chi}$, $\delta_{\theta}$, and $\delta_{\chi}$ predicted by the associated smoothing splines to fully describe the polarimetric response of the instrument at any epoch and radio frequency spanned by the splines.

The critical steps detailed above are outlined here:
\begin{enumerate}
    \item Prepare Data
    \begin{itemize}
        \item Collect observations of the polarized reference source and its calibrators to create MEM sessions, ensuring that each session has good parallactic angle coverage and sufficient signal-to-noise ratio for accurate analysis. 
    \end{itemize}
    
    \item Produce Calibrator Models
    \begin{itemize}
        \item Produce MEM calibrator model solutions.
        \item Create template archives for METM analysis.
        \item Produce METM calibrator model solutions.
        \item Correct ionospheric Faraday rotation.
        \item Produce spline-smoothed METM solutions.
    \end{itemize}
    
    \item Calibrate Pulsar Data
    \begin{itemize}
        \item Apply calibration solutions to pulsar observations.
    \end{itemize}
\end{enumerate}
After completing these steps, arrival times are computed as described in the following section. A more detailed outline of the  \textsc{PSRCHIVE} commands used for both polarimetric calibration and arrival time estimation is provided in Appendix~\ref{app:steps}.

\subsection{Arrival Time Estimation and Analysis}\label{sec:timing}
Arrival times are estimated using both conventional \emph{Scalar Template Matching} \citep[STM;][]{Taylor_1992} and \emph{Matrix Template Matching} \citep[MTM;][]{Straten_2006}. MTM quadruples the number of observational constraints while introducing only six degrees of freedom. For 23 out of 25 pulsars in our analysis, arrival time estimates derived from the polarization profile (using MTM) are expected to have greater precision than those derived from the total intensity profile alone (using STM), as indicated by the theoretical relative ToA uncertainty between MTM and STM ($\hat{\sigma}_{\varphi}$ in Table~\ref{tab:timing_predictions}). 
\begin{table}
\begin{center}
    \caption{{Relative Arrival Time Uncertainties for Each Pulsar. Columns, from left to right, include J2000.0 coordinates (JNAME), predicted timing error for a 1\% calibration error ($\tau_{\beta}$), {PPTA DR2} timing precision \citep[$\sigma_{\tau}$,][]{Kerr_2020}, fraction of {PPTA DR2} timing residuals that could be due to predicted calibration error \citep[$\tau_{\beta}/\sigma_{\tau}$,][]{Straten_2013}, theoretical relative ToA uncertainty between MTM and STM template matching algorithms ($\hat{\sigma}_{\varphi}$), and the ratio between uncertainties in arrival times derived from the invariant interval and total intensity ($\hat{\sigma}_{\tilde{\varphi}}$). For PSR~J0437$-$4715, {PPTA DR2 timing precision} is acheived by timing the invariant profile \citep{Britton_2000}.}}
    \label{tab:timing_predictions}
    
    \begin{tabular}{c c c c c c}
        \toprule
        Pulsar & $\tau_{\beta}$ & $\sigma_{\tau}$ & $\tau_{\beta}/\sigma_{\tau}$ & $\hat{\sigma}_{\varphi}$ & $\hat{\sigma}_{\tilde{\varphi}}$ \\
        (JNAME) & (ns) & (ns) & & & \\
        \midrule
        PSR~J0437$-$4715  & 205 &  116 & 1.77  & 0.82 & 1.43 \\
        PSR~J0613$-$0200  &  59 & 1018 & 0.06  & 0.95 & 1.49 \\
        PSR~J0711$-$6830  &  73 &  979 & 0.07  & 0.89 & 1.53 \\
        PSR~J1017$-$7156  &  74 &  635 & 0.12  & 0.92 & 1.58 \\
        PSR~J1022+1001    & 278 & 1555 & 0.18  & 0.73 & 1.67 \\
        PSR~J1024$-$0719  &  33 & 1023 & 0.03  & 0.73 & 2.20 \\
        PSR~J1045$-$4509  & 338 & 2570 & 0.13  & 0.87 & 1.50 \\
        PSR~J1125$-$6014  &   8 & 1510 & 0.005 & 0.94 & 1.30 \\ 
        PSR~J1446$-$4701  &  85 & 1359 & 0.08  & 0.96 & 1.33 \\
        PSR~J1545$-$4550  &  66 &  808 & 0.08  & 0.84 & 1.74 \\      
        PSR~J1600$-$3053  & 119 &  572 & 0.21  & 0.88 & 1.39 \\
        PSR~J1603$-$7202  & 143 & 1316 & 0.11  & 0.84 & 1.56 \\
        PSR~J1643$-$1224  & 269 & 2248 & 0.12  & 0.92 & 1.39 \\
        PSR~J1713+0747    &   5 &  287 & 0.02  & 0.86 & 1.57 \\
        PSR~J1730$-$2304  & 196 & 1322 & 0.15  & 0.74 & 1.69 \\
        PSR~J1744$-$1134  & 108 &  385 & 0.28  & 1.59 & 6.98 \\
        PSR~J1824$-$2452A &  20 & 2628 & 0.008 & 0.84 & 4.01 \\
        PSR~J1832$-$0836  &  17 &  563 & 0.03  & 0.95 & 1.44 \\
        PSR~J1857+0943    & 121 & 1208 & 0.10  & 0.92 & 1.43 \\
        PSR~J1909$-$3744  &  22 &  152 & 0.15  & 0.83 & 1.52 \\
        PSR~J1939+2134    &  44 &  586 & 0.08  & 0.92 & 1.49 \\
        PSR~J2124$-$3358  & 115 & 2551 & 0.05  & 0.85 & 1.45 \\
        PSR~J2129$-$5721  & 225 &  964 & 0.23  & 1.08 & 1.61 \\
        PSR~J2145$-$0750  & 147 &  995 & 0.15  & 0.95 & 1.45 \\
        PSR~J2241$-$5236  &  21 &  334 & 0.06  & 0.98 & 1.36 \\
        \bottomrule
    \end{tabular}
\end{center}
\end{table}
%
Table~\ref{tab:timing_predictions} also shows the predicted timing error for a 1\% calibration error \citep[$\tau_{\beta}$;][]{Straten_2013} and the timing precision achieved by {PPTA DR2} \citep[$\sigma_{\tau}$;][]{Kerr_2020}. For a given MSP, the potential significance of systematic timing errors due to inaccurate instrumental calibration is characterised by $\tau_{\beta}/\sigma_{\tau}$ (column 4).

The predicted values of relative uncertainty $\hat{\sigma}_{\varphi}$ are based on analysis of spectral content in the phase-resolved average profiles of all four Stokes parameters and additional (white) radiometer noise. However, over sufficiently long timescales, pulse arrival time residuals exhibit (red) timing noise, and it is necessary to model and remove this red noise before the white noise content of residuals can be quantified and compared. In principle, the systematic timing errors induced by polarization distortions could also induce red noise, such as the $1/f$ noise produced by unmodeled steps in instrumental response. Therefore, it is also interesting to quantify and compare the red noise content of pulsar timing residuals. 

In this study, we analyze both the white and red noise components of timing residuals using \textsc{Tempo2} \citep{Edwards_2006, Hobbs_2006}\footnote{https://www.atnf.csiro.au/research/pulsar/tempo2/} 
and \textsc{TempoNest} \citep{Lentati_2014}\footnote{https://github.com/LindleyLentati/TempoNest}.
\textsc{Tempo2} is employed to fit the timing model to the observed ToAs by minimising the timing residuals. \textsc{TempoNest} utilizes the multi-modal nested sampling algorithm \textsc{MultiNest} \citep{Feroz_2009}\footnote{https://github.com/farhanferoz/MultiNest} to explore the parameter space of the non-linear pulsar timing model. Simultaneously, it determines a red noise model and two time-independent white noise modifiers:
\begin{itemize}
    \item Error Scale Factor (EFAC): This accounts for any errors that are proportional to the estimated uncertainty (including potential miscalibrated radiometer noise in the system). It modifies each ToA uncertainty by a constant scale factor $E_{f}$.
    \item Error Added in Quadrature (EQUAD): This compensates for additional white noise by adding a constant $E_q$ in quadrature to each ToA uncertainty \citep{Lentati_2014}. 
\end{itemize}
The total ToA uncertainty is therefore obtained by adjusting the uncertainty $\sigma_\tau$ as follows:
\begin{equation}
\sigma_\tau' = \sqrt{E_q^{2} + E_{f}^{2}\sigma{^{2}_{\tau}}}
\end{equation}
EFAC and EQUAD values are typically applied to all TOAs in a pulsar timing data set and adjusted iteratively until the fitted model's reduced $\chi^{2}$ reaches unity \citep{Shannon_2014}.

\textsc{TempoNest} models red noise using a power-law spectrum characterized by an amplitude $(A_\mathrm{red})$ and spectral index $\beta$:
\begin{equation}
P_{r}(f) = \frac{A_\mathrm{red}^{2}}{12\pi^{2}} \left(\frac{f}{f_\mathrm{yr}}\right)^{-\beta}
\label{eqn:red_noise_model}
\end{equation}
Here, $f_\mathrm{yr}$ is a reference frequency of 1 cycle per year and the amplitude $A_\mathrm{red}$ is in units of $\mathrm{yr}^{3/2}$
%
%
%
\citep{Lentati_2014, Parthasarathy_2019}. The prior ranges for the red noise and additional white noise parameters used in our analysis are detailed in Table~\ref{tab:prior_ranges}. Note that \textsc{TempoNest} is configured to sample the spectrum using 120 Fourier coefficients.
%
%
\begin{table}
    \centering
    \caption{Prior ranges on white and red noise model parameters, {comprising the dimensionless base-10 logarithms of the error scale factor $\hat{E}_{f,10}=\log_{10}(E_{f})$, the error added in quadrature $\hat{E}_{q,10}=\log_{10}(E_q/\mathrm{s})$, and the red-noise amplitude $\hat{A}_\mathrm{red,10}=\log_{10}(A_\mathrm{red}/\mathrm{yr}^{3/2})$; and the power-law spectral index $\beta$.}}
    \label{tab:prior_ranges}
    \begin{tabular}{c c c}
        \toprule
        Parameter & Prior Range & Type\\
        \midrule
        $\hat{E}_{f,10}=\log_{10}(E_{f})$ & (-1,1) & log-uniform\\
        $\hat{E}_{q,10}=\log_{10}(E_q/\mathrm{s})$ & (-9,-5) & log-uniform\\
        $\hat{A}_\mathrm{red,10}=\log_{10}(A_\mathrm{red}/\mathrm{yr}^{3/2})$ & (-18,-10) & log-uniform\\
        $\beta$ & (0,7) & log-uniform\\
        \bottomrule
    \end{tabular}
\end{table}

Outliers in pulsar timing measurements bias both timing model and noise model parameter estimates and reduce the accuracy of estimated parameter uncertainties \citep{Vallisneri_2017}. To address this challenge, we employ robust and automated outlier removal techniques. An initial outlier rejection involves removing any instances of ToAs with error equal to 0 and performing a 5-sigma outlier removal on both the relative error for each residual (residual divided by uncertainty) and ToA goodness-of-fit. We then run \textsc{Tempo2} with the initial (slightly corrupted) \textsc{TempoNest} red-noise model, producing whitened residuals that are corrupted by some remaining outliers. These remaining outliers are removed using Tukey's Fence \citep{Tukey_1977}, a robust statistical method for outlier detection and removal \citep{Morello_2019}, which is used to enhance timing accuracy. This approach typically flags only a small portion (around 5\%) of the data as outliers and has been shown to improve timing accuracy by a factor of two for many pulsars \citep{Lower_2020}. Tukey's Fence defines a "reasonable" range based on the interquartile range (IQR), which is the difference between the 75$^{th}$ percentile (Q${3}$) and the 25$^{th}$ percentile (Q${1}$) of the data set. The range extends from Q${1}$ - $q$*IQR to Q${3}$ + $q$*IQR, where $q$ is a parameter. While a common choice for $q$ is 1.5, it can be adjusted according to specific analysis requirements, determining the stringency of the outlier cutoff. We used a less stringent cutoff value of $q$=2 for our analysis.

Following the application of Tukey's Fence to each MSP, we update the best-fit model by re-running \textsc{TempoNest} to generate a red noise model that is not corrupted by outliers. PSR~J0437$-$4715 and PSR~J2241$-$5721 are exceptions. An additional 5-sigma removal was performed on the relative error for each residual to catch a few outliers missed by Tukey's Fence for these pulsars. A final run of \textsc{Tempo2} with the final TempoNest red-noise model produces whitened residuals for each calibration and ToA estimation method combination. By comparing the red and white noise model parameters, we aim to evaluate whether advanced techniques like METM and MTM, which better account for instrumental calibration errors, can enhance the experimental sensitivity of PTAs over extended periods compared to conventional methods like IFA and STM.

\section{RESULTS}\label{sec:RESULTS}

\begin{figure*}
\centering
\includegraphics[width=\textwidth]{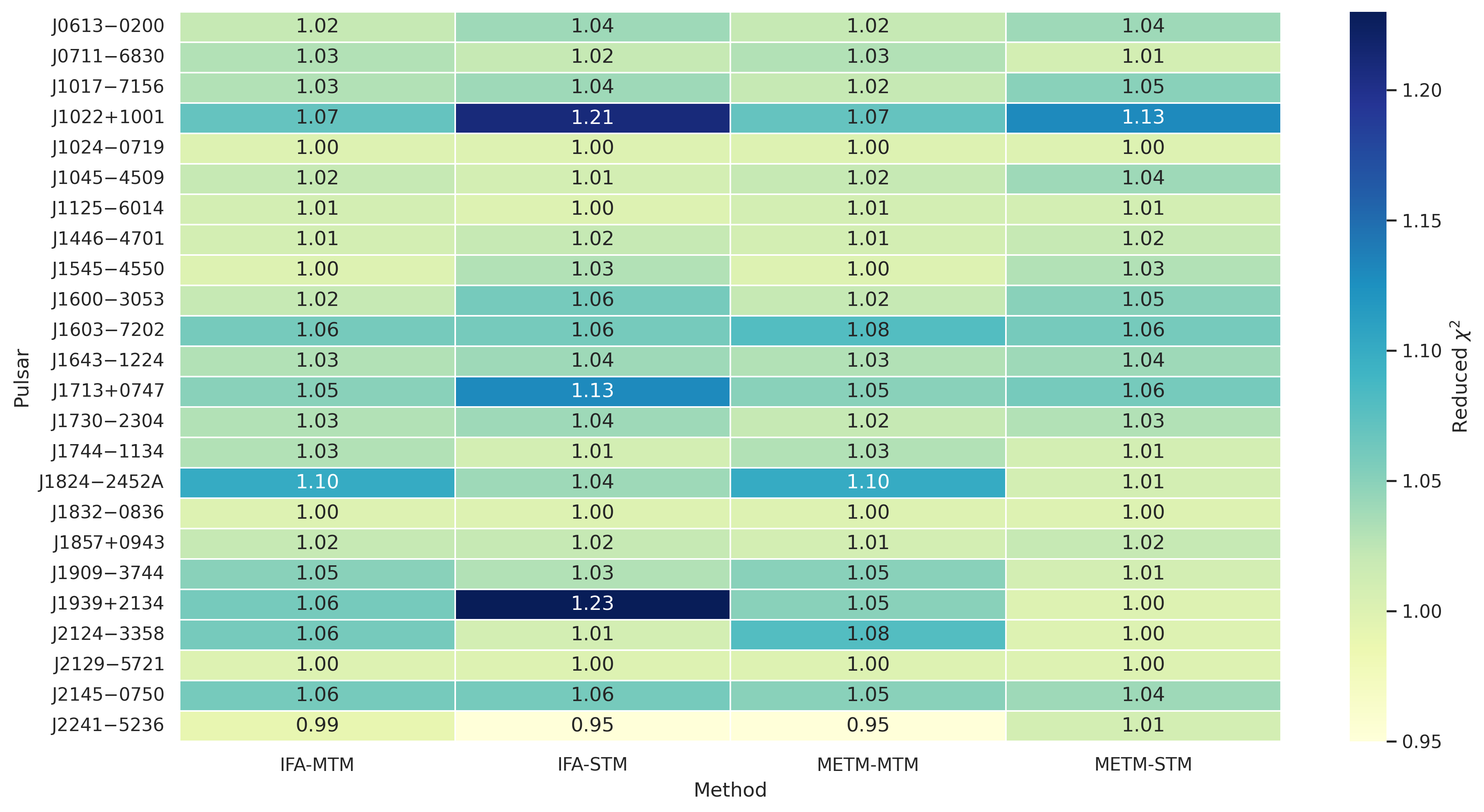}
\caption{The reduced $\chi^{2}$ heat map illustrates the median goodness-of-fit over all arrival times obtained for each pulsar. PSR~J0437$-$4715 is excluded owing to exceptionally large values of 2.34, 7.77 2.04, and 2.61 for the IFA-MTM, IFA-STM, METM-MTM, and METM-STM methods, respectively.}
\label{fig:heatmap}
\end{figure*}

In this section, we present the results of our analysis, comparing advanced polarimetric calibration \citep[METM;][]{Straten_2013} and arrival time estimation \citep[MTM;][]{Straten_2006} methods with conventional approaches (IFA and STM). We applied these techniques to 25 high-priority pulsars regularly observed as part of the PPTA project, resulting in four distinct data sets (IFA-MTM, METM-MTM, IFA-STM, and METM-STM).

The quality of the arrival time estimation procedure is summarised in Figure~\ref{fig:heatmap}, which presents the median goodness-of-fit for each pulsar, as characterised by the reduced $\chi^{2}$ of the fit between the observed pulse profile and the template pulse profile.
Large values of the reduced $\chi^{2}$ may indicate that the template pulse profile is not an accurate representation of the 
average pulse profile, or that the observed pulse profiles vary by more than what would be expected due to radiometer noise alone; e.g. owing to additional pulsar self-noise known as jitter \citep{Cordes_1985, Oslowski_2011}.

The results of modeling the arrival time estimates 
are summarised in Table~\ref{tab:timing_results}.

\startlongtable
\begin{deluxetable*}{ccccccccc}
\tablecaption{Noise statistics and noise model parameters for each timing data set. 
Columns, from left to right, include each pulsar's J2000.0 coordinates; the calibration-ToA estimation method; the number of TOAs, $N_\mathrm{toas}$; the uncertainty-weighted standard deviation of 
the post-fit timing residuals $\sigma_\tau$, and whitened timing residuals \rmswhite; 
and the maximum likelihood estimates of the noise model parameters, comprising the
{dimensionless base-10 logarithms of the} 
error scale factor $\hat{E}_{f,10}=\log_{10}(E_{f})$, 
the error added in quadrature $\hat{E}_{q,10}=\log_{10}(E_q/\mathrm{s})$, 
and the red-noise amplitude $\hat{A}_\mathrm{red,10}=\log_{10}(A_\mathrm{red}/\mathrm{yr}^{3/2})$; 
and the power-law spectral index $\beta$. 
Values in parentheses are the $1\sigma$ uncertainty in the last digit quoted. \label{tab:timing_results}}
\tablehead{
        \colhead{Pulsar}
        & \colhead{Method}
        & \colhead{$N_\mathrm{toas}$}
        & \colhead{$\dfrac{\sigma_\tau}{\mu\mathrm{s}}$}
        & \colhead{$\dfrac{\rmswhite}{\mu\mathrm{s}}$}
        & \colhead{$\hat{E}_{f,10}$}
        & \colhead{$\hat{E}_{q,10}$}
        & \colhead{$\hat{A}_\mathrm{red,10}$}
        & \colhead{$\beta$}\\[-5mm]}
\startdata
PSR~J0437$-$4715  
& METM-MTM & 700 & 0.584 & \textcolor{blue}{0.100}  &  0.60(5) & -7.02(2) & -13.48(7) & 3.0(4) \\
& IFA-MTM  & 700 & 0.653 & 0.103                    &  0.64(5) & -7.02(2) & -13.47(7) & 3.1(4) \\
& METM-STM & 700 & 0.680 & 0.231                    &  0.6(1)  & -6.64(1) & -13.5(1)  & 3.1(7) \\
& IFA-STM  & 700 & 0.947 & 0.663                    &  0.3(4)  & -6.17(1) & -13.1(1)  & 2.0(5) \\ 
\tablebreak
PSR~J0613$-$0200
& METM-MTM & 221 & 0.661 & \textcolor{blue}{0.583}  &  0.04(3) & -7.0(4)  & -15(1)    & 4(2) \\
& IFA-MTM  & 221 & 0.648 & \textcolor{blue}{0.583}  &  0.04(3) & -7.0(4)  & -16(1)    & 4(2) \\
& METM-STM & 221 & 1.111 & 1.083                    &  0.22(6) & -6.3(3)  & -16(1)    & 3(2) \\ 
& IFA-STM  & 221 & 1.200 & 1.117                    &  0.22(7) & -6.3(3)  & -16(1)    & 3(2) \\
\midrule
PSR~J0711$-$6830  
& METM-MTM & 333 & 0.816 & 0.754                    & -0.02(2) & -6.6(3)  & -14(1)    & 2(2) \\
& IFA-MTM  & 333 & 0.805 & \textcolor{blue}{0.722}  & -0.01(2) & -6.8(3)  & -13.8(8)  & 2(2) \\
& METM-STM & 333 & 0.925 & 0.842                    &  0.02(2) & -6.7(4)  & -13.6(9)  & 2(2) \\
& IFA-STM  & 333 & 1.069 & 0.972                    &  0.04(2) & -6.6(3)  & -14.0(7)  & 4(2) \\
\midrule
PSR~J1017$-$7156
& METM-MTM & 333 & 1.704 & 0.223                    & -0.09(2) & -7.4(4)  & -12.88(6) & 2.7(4) \\
& IFA-MTM  & 333 & 1.726 & \textcolor{blue}{0.221}  & -0.09(2) & -7.4(4)  & -12.89(5) & 2.7(4) \\
& METM-STM & 333 & 1.751 & 0.294                    &  0.14(2) & -7.3(4)  & -12.90(6) & 2.9(5) \\
& IFA-STM  & 333 & 1.800 & 0.312                    &  0.15(2) & -7.2(4)  & -12.90(6) & 2.9(5) \\
\midrule
PSR~J1022+1001
& METM-MTM & 215 & 0.857 & \textcolor{blue}{0.833}  &  0.00(5) & -6.19(4) & -16(1)    & 3(2) \\
& IFA-MTM  & 215 & 0.875 & 0.864                    &  0.01(5) & -6.17(4) & -16(1)    & 3(2) \\
& METM-STM & 215 & 1.341 & 1.281                    &  0.07(4) & -6.00(4) & -15(1)    & 3(2) \\
& IFA-STM  & 215 & 1.619 & 1.619                    &  0.04(5) & -5.88(4) & -16(1)    & 3(2) \\
\midrule
PSR~J1024$-$0719  
& METM-MTM & 127 & 0.739 & \textcolor{blue}{0.654}  &  0.05(3) & -6.9(4)  & -15(1)    & 4(2) \\
& IFA-MTM  & 127 & 0.787 & 0.721                    &  0.07(3) & -6.8(4)  & -15(1)    & 4(2) \\
& METM-STM & 127 & 1.088 & 1.078                    &  0.11(3) & -6.9(5)  & -16(1)    & 3(2) \\
& IFA-STM  & 127 & 1.095 & 1.081                    &  0.12(3) & -6.9(5)  & -16(1)    & 3(2) \\
\midrule
PSR~J1045$-$4509
& METM-MTM & 177 & 4.025 & \textcolor{blue}{1.280}  & -0.07(5) & -6.5(5)  & -12.46(9) & 2.6(5) \\
& IFA-MTM  & 177 & 4.689 & 1.311                    & -0.06(4) & -6.5(4)  & -12.44(9) & 2.6(5) \\
& METM-STM & 177 & 7.010 & 4.364                    & -0.3(3)  & -5.34(4) & -12.5(2)  & 2.7(9) \\
& IFA-STM  & 177 & 5.739 & 4.702                    &  0.3(1)  & -5.4(1)  & -12.5(4)  & 2(1)   \\
\midrule
PSR~J1125$-$6014
& METM-MTM & 141 & 1.939 & \textcolor{blue}{0.833}  &  0.0(1)  & -6.06(5) & -12.9(1)  & 2.8(7) \\
& IFA-MTM  & 141 & 1.919 & 0.836                    &  0.0(1)  & -6.07(5) & -12.8(1)  & 2.6(6) \\
& METM-STM & 141 & 2.030 & 0.844                    &  0.0(1)  & -6.08(5) & -12.8(1)  & 2.8(7) \\
& IFA-STM  & 141 & 1.993 & \textcolor{blue}{0.833}  &  0.0(1)  & -6.08(5) & -12.8(1)  & 2.7(6) \\
\midrule
PSR~J1446$-$4701
& METM-MTM & 145 & 1.747 & \textcolor{blue}{1.318}  & -0.01(7) & -6.00(9) & -15(1)    & 3(2) \\
& IFA-MTM  & 145 & 1.605 & 1.345                    &  0.04(6) & -6.0(1)  & -15(1)    & 3(2) \\
& METM-STM & 145 & 2.132 & 1.978                    &  0.34(7) & -5.8(1)  & -16(1)    & 3(2) \\
& IFA-STM  & 145 & 2.247 & 2.227                    &  0.33(8) & -5.8(1)  & -16(1)    & 3(2) \\
\midrule
PSR~J1545$-$4550	
& METM-MTM & 123 & 0.594 & \textcolor{blue}{0.429}  & -0.1(1)  & -6.9(4)  & -13.3(2)  & 3(1) \\
& IFA-MTM  & 123 & 0.603 & 0.434                    &  0.06(7) & -6.9(4)  & -13.3(2)  & 3(1) \\
& METM-STM & 123 & 1.068 & 0.776                    &  0.13(9) & -6.6(4)  & -13.2(3)  & 3(1) \\
& IFA-STM  & 123 & 1.052 & 0.785                    &  0.12(9) & -6.5(4)  & -13.1(3)  & 3(1) \\
\tablebreak
PSR~J1600$-$3053
& METM-MTM & 174 & 1.828 & \textcolor{blue}{0.230}  & -0.04(6) & -7.0(4)  & -13.3(2)  & 2.9(6) \\
& IFA-MTM  & 174 & 1.748 & 0.221                    &  0.05(5) & -7.2(4)  & -13.2(1)  & 2.7(5) \\
& METM-STM & 174 & 1.757 & 0.311                    &  0.08(6) & -7.1(4)  & -14.0(5)  & 5(1)   \\
& IFA-STM  & 174 & 2.100 & 0.341                    &  0.10(7) & -6.9(3)  & -14.1(5)  & 5(1)   \\
\midrule
PSR~J1603$-$3053
& METM-MTM & 212 & 0.783 & \textcolor{blue}{0.497}  &  0.01(3) & -7.2(4)  & -13.3(2)  & 2.6(8) \\
& IFA-MTM  & 212 & 0.819 & 0.504                    &  0.01(3) & -7.2(4)  & -13.3(2)  & 2.8(9) \\
& METM-STM & 212 & 0.877 & 0.593                    &  0.00(3) & -7.1(4)  & -13.2(2)  & 2.6(8) \\
& IFA-STM  & 212 & 0.961 & 0.747                    &  0.06(3) & -6.8(4)  & -13.4(3)  & 3(1)   \\
\midrule
PSR~J1643$-$1224
& METM-MTM & 155 & 2.481 & 0.452                    & -0.12(6) & -6.8(4)  & -12.65(7)  & 2.0(3) \\
& IFA-MTM  & 155 & 2.153 & \textcolor{blue}{0.438}  & -0.14(7) & -6.7(4)  & -12.64(6)  & 1.9(3) \\
& METM-STM & 155 & 2.282 & 1.087                    &  0.26(4) & -6.6(4)  & -12.8(2)   & 2.6(7) \\
& IFA-STM  & 155 & 2.133 & 1.039                    &  0.24(4) & -6.6(4)  & -12.8(2)   & 2.4(7) \\
\midrule
PSR~J1713+0747
& METM-MTM & 246 & 0.234 & \textcolor{blue}{0.201}  & -0.08(5) & -6.76(3) & -13.6(1) & 1.5(6) \\
& IFA-MTM  & 246 & 0.232 & 0.204                    & -0.07(6) & -6.75(3) & -13.6(1) & 1.5(6) \\
& METM-STM & 246 & 0.252 & 0.216                    & -0.04(6) & -6.72(3) & -13.6(1) & 1.6(6) \\
& IFA-STM  & 246 & 0.247 & 0.215                    & -0.01(6) & -6.72(3) & -13.6(1) & 1.5(6) \\        
\midrule
PSR~J1730$-$2304
& METM-MTM & 165 & 0.935 & 0.668                    &  0.11(3) & -7.1(4)  & -13.7(4) & 3(1) \\
& IFA-MTM  & 165 & 0.865 & \textcolor{blue}{0.664}  &  0.11(3) & -7.2(4)  & -13.7(4) & 3(1) \\
& METM-STM & 165 & 1.460 & 0.963                    &  0.11(3) & -6.9(5)  & -13.5(4) & 3(1) \\
& IFA-STM  & 165 & 1.488 & 1.171                    &  0.14(4) & -6.4(4)  & -13.7(8) & 4(1) \\
\midrule
PSR~J1744$-$1134
& METM-MTM & 237 & 0.426 & 0.306                    &  0.02(3) & -6.8(2)  & -13.5(4) & 2(1) \\
& IFA-MTM  & 237 & 0.438 & 0.311                    &  0.04(3) & -6.9(2)  & -13.5(4) & 2(1) \\
& METM-STM & 237 & 0.442 & \textcolor{blue}{0.301}  & -0.09(4) & -6.63(5) & -13.6(4) & 2(1) \\
& IFA-STM  & 237 & 1.034 & 0.682                    & -0.13(6) & -6.21(3) & -14.5(6) & 5(2) \\
\midrule
PSR~J1824$-$2452A
& METM-MTM & 80 & 16.467 & 0.537                    &  0.1(4)  & -6.2(2)  & -12.4(1) & 3.5(6) \\
& IFA-MTM  & 80 & 15.780 & \textcolor{blue}{0.507}  &  0.0(4)  & -6.2(2)  & -12.3(1) & 3.3(7) \\
& METM-STM & 80 & 16.445 & 0.551                    &  0.2(3)  & -6.3(2)  & -12.4(1) & 3.6(6) \\
& IFA-STM  & 80 & 16.148 & 0.510                    & -0.1(4)  & -6.2(2)  & -12.4(1) & 3.5(7) \\
\midrule
PSR~J1832$-$0836
& METM-MTM & 72 & 2.485 & \textcolor{blue}{0.597}   & -0.02(6) & -6.8(4)  & -14.1(4) & 6(1) \\
& IFA-MTM  & 72 & 2.066 & 0.645                     &  0.03(5) & -6.8(4)  & -14.0(4) & 5(1) \\
& METM-STM & 72 & 1.605 & 0.929                     &  0.1(1)  & -6.4(4)  & -13.9(5) & 5(1) \\
& IFA-STM  & 72 & 1.593 & 0.809                     &  0.0(2)  & -6.3(4)  & -13.0(5) & 3(2) \\
\midrule
PSR~J1857+0943
& METM-MTM & 127 & 1.288 & \textcolor{blue}{0.573}  & -0.06(5) & -6.8(4)  & -13.9(4) & 4(1) \\
& IFA-MTM  & 127 & 1.255 & 0.597                    & -0.04(5) & -6.8(4)  & -13.8(4) & 4(1) \\
& METM-STM & 127 & 1.335 & 0.849                    &  0.02(6) & -6.6(4)  & -13.8(4) & 4(1) \\
& IFA-STM  & 127 & 1.532 & 0.889                    &  0.04(6) & -6.5(4)  & -13.9(4) & 4(1) \\
\tablebreak
PSR~J1909$-$3744
& METM-MTM & 344 & 0.518 & \textcolor{blue}{0.200}  & -0.10(6) & -6.73(3) & -14.0(2) & 3.7(8) \\
& IFA-MTM  & 344 & 0.512 & 0.204                    & -0.11(7) & -6.72(3) & -14.0(2) & 3.7(8) \\
& METM-STM & 344 & 0.530 & 0.204                    & -0.08(6) & -6.72(3) & -14.0(3) & 3.6(9) \\
& IFA-STM  & 344 & 0.540 & 0.204                    & -0.13(7) & -6.72(3) & -14.0(3) & 3.8(9) \\
\midrule
PSR~J1939+2134
& METM-MTM & 142 & 1.121 & \textcolor{blue}{0.101}  &  0.4(2)  & -7.0(2)  & -12.94(6) & 2.5(3) \\
& IFA-MTM  & 142 & 1.126 & \textcolor{blue}{0.101}  &  0.4(2)  & -7.0(2)  & -12.95(6) & 2.5(3) \\
& METM-STM & 142 & 1.117 & 0.104                    &  0.1(3)  & -6.93(7) & -12.98(6) & 2.7(3) \\
& IFA-STM  & 142 & 1.173 & 0.104                    &  0.6(1)  & -7.3(3)  & -12.96(6) & 2.5(3) \\
\midrule
PSR~J2124$-$3358
& METM-MTM & 221 & 1.341 & 1.320                    & -0.01(2) & -6.9(5)  & -16(1)    & 3(2) \\
& IFA-MTM  & 221 & 1.275 & \textcolor{blue}{1.272}  & -0.02(2) & -6.9(5)  & -16(1)    & 3(2) \\
& METM-STM & 221 & 2.936 & 2.983                    &  0.37(2) & -6.4(5)  & -16(1)    & 3(2) \\
& IFA-STM  & 221 & 2.920 & 2.886                    &  0.38(3) & -6.4(5)  & -16(1)    & 3(2) \\
\midrule
PSR~J2129$-$5721
& METM-MTM & 238 & 1.131 & 0.914                    &  0.01(3) & -6.7(4)  & -14.7(6)  & 5(1) \\
& IFA-MTM  & 238 & 1.162 & \textcolor{blue}{0.911}  &  0.00(3) & -6.7(4)  & -14.4(6)  & 5(1) \\
& METM-STM & 238 & 1.301 & 1.151                    &  0.17(2) & -6.9(4)  & -15(1)    & 4(2) \\
& IFA-STM  & 238 & 1.336 & 1.194                    &  0.17(2) & -6.9(4)  & -14.4(7)  & 5(2) \\
\midrule
PSR~J2145$-$0750
& METM-MTM & 211 & 0.592 & \textcolor{blue}{0.531}  &  0.00(3) & -6.51(7) & -16(1)    & 4(2) \\
& IFA-MTM  & 211 & 0.581 & \textcolor{blue}{0.531}  &  0.02(3) & -6.51(7) & -15(1)    & 4(2) \\
& METM-STM & 211 & 0.832 & 0.602                    &  0.00(3) & -6.41(6) & -14.6(8)  & 4(2) \\
& IFA-STM  & 211 & 0.810 & 0.684                    &  0.03(4) & -6.32(7) & -15(1)    & 4(2) \\
\midrule
PSR~J2241$-$5236
& METM-MTM & 329 & 0.251 & \textcolor{blue}{0.214}  &  0.14(4) & -6.80(6) & -13.9(4)  & 3(1) \\
& IFA-MTM  & 329 & 0.255 & 0.217                    &  0.13(5) & -6.80(6) & -14.0(5)  & 3(1) \\
& METM-STM & 329 & 0.254 & 0.215                    &  0.15(3) & -6.87(6) & -14.0(3)  & 3(1) \\
& IFA-STM  & 329 & 0.269 & 0.229                    &  0.14(4) & -6.80(5) & -13.9(3)  & 3(1) \\
\enddata
\end{deluxetable*}

For each pulsar and each combination of methods, Table~\ref{tab:timing_results}
lists the number of ToAs, the weighted standard deviations of the post-fit residuals output by \textsc{Tempo2} and the noise model parameters obtained through \textsc{TempoNest}.
The parameters that characterise the white noise of each pulsar are summarised in Table~\ref{tab:error_corrections}, which compares the minimum, median, and maximum values of $E_f$, $E_q$, and the uncertainty-weighted standard deviation of the whitened (red noise removed) post-fit timing residuals, \rmswhite, for the four data sets. Notably, IFA-STM has the largest median values of $E_f$ and $E_q$; therefore, it serves as the baseline for model comparison in this study.

\begin{table*}
\caption{{Minimum, median, and maximum values of the} error scale factor $E_{f}$, error added in quadrature $E_q$, and uncertainty-weighted standard deviation of the whitened post-fit timing residuals \rmswhite, for each combination of calibration and arrival time estimation methods.}
\begin{center}    
\begin{tabular}{c|c c c|c c c|c c c}
\toprule
\multirow{2}{*}{Model} & \multicolumn{3}{c|}{$E_{f}$} & \multicolumn{3}{c|}{$E_q$ ($\mu$s)} & \multicolumn{3}{c}{$\rmswhite$ ($\mu$s)}\\
 & Min & Med & Max & Min & Med & Max & Min & Med & Max \\
\midrule
METM-MTM & 0.77 & 1.02 & 4.02 & 0.04 & 0.16 & 0.99 & 0.10 & 0.54 & 1.32 \\
IFA-MTM  & 0.72 & 1.07 & 4.34 & 0.04 & 0.15 & 0.95 & 0.10 & 0.51 & 1.35 \\
METM-STM & 0.53 & 1.27 & 4.21 & 0.05 & 0.23 & 4.61 & 0.10 & 0.84 & 4.36 \\
IFA-STM  & 0.75 & 1.30 & 4.13 & 0.05 & 0.33 & 3.85 & 0.10 & 0.81 & 4.70 \\
\bottomrule
\end{tabular}
\end{center}
\label{tab:error_corrections}
\end{table*}
Table~\ref{tab:significant_equad_differences} compares the best-fit $E_q$ estimates of the IFA-STM and METM-MTM data sets for those {seven} pulsars with statistically significant differences.
\begin{table}
\centering
\caption{Pulsars with {statistically significant} different estimates of error added in quadrature, $E_q$.
The best-fit estimates of $E_q$ derived from the IFA-STM and METM-MTM data sets are shown in
columns 2 and 3, and the {quadrature} differences between them are listed in column 4.
Values in parentheses are the $1\sigma$ uncertainty in the last digit quoted.}
\begin{tabular}{c c c c c c c c c}
\toprule
Pulsar            & \multicolumn{2}{c}{$E_q$ ($\mu$s) } & $\Delta E_q$ ($\mu$s) \\
(JNAME)           & IFA-STM  & METM-MTM   &         \\
\midrule
PSR~J0437$-$4715  & 0.67(2)  &  0.095(4)  & 0.67(2) \\
PSR~J1022+1001    & 1.3(1)   &  0.64(7)   & 1.1(1)  \\
PSR~J1045$-$4509  & 4(1)     &  0.3(3)    & 4(1)    \\
PSR~J1446$-$4701  & 1.7(4)   &  1.0(2)    & 1.4(4)  \\
PSR~J1713+0747    & 0.19(1)  &  0.17(1)   & 0.08(2) \\
PSR~J1744$-$1134  & 0.62(4)  &  0.15(6)   & 0.60(7) \\
PSR~J2145$-$0750  & 0.48(8)  &  0.31(5)   & 0.37(9) \\
\bottomrule
\end{tabular}
\label{tab:significant_equad_differences}
\end{table}

\begin{figure*}
\centering
\gridline{\fig{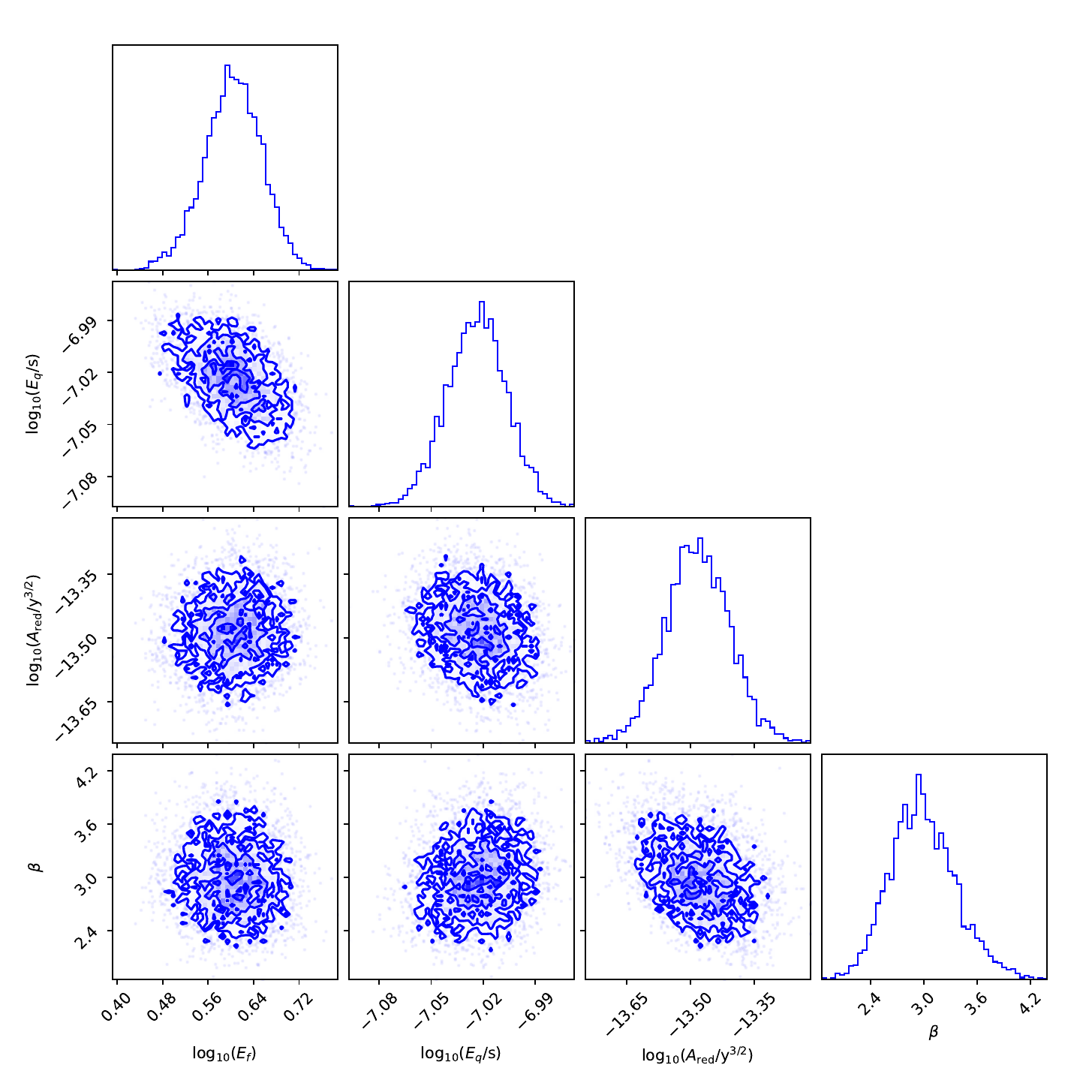}{0.48\textwidth}{(a) METM-MTM}
          \fig{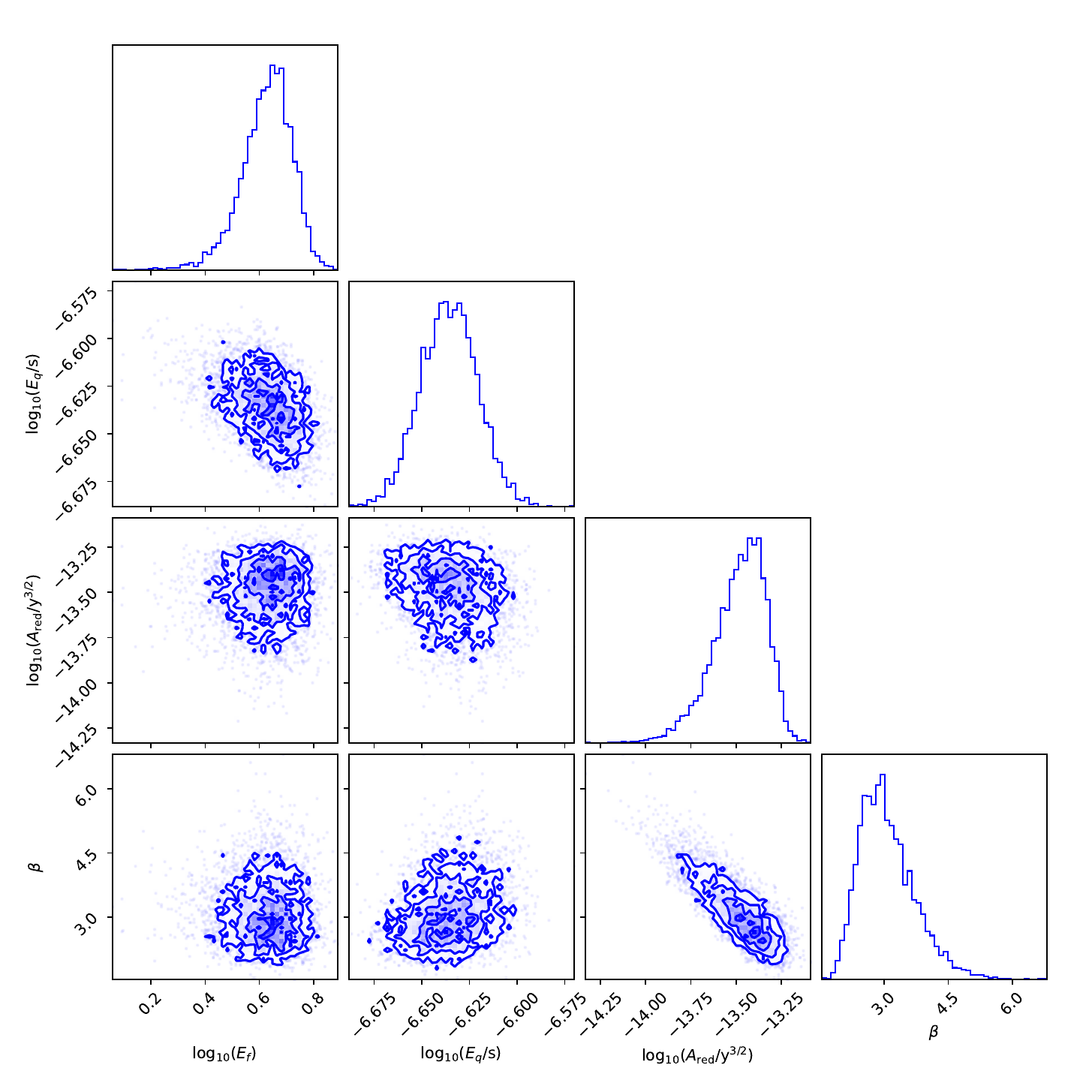}{0.48\textwidth}{(b) METM-STM}}
\gridline{\fig{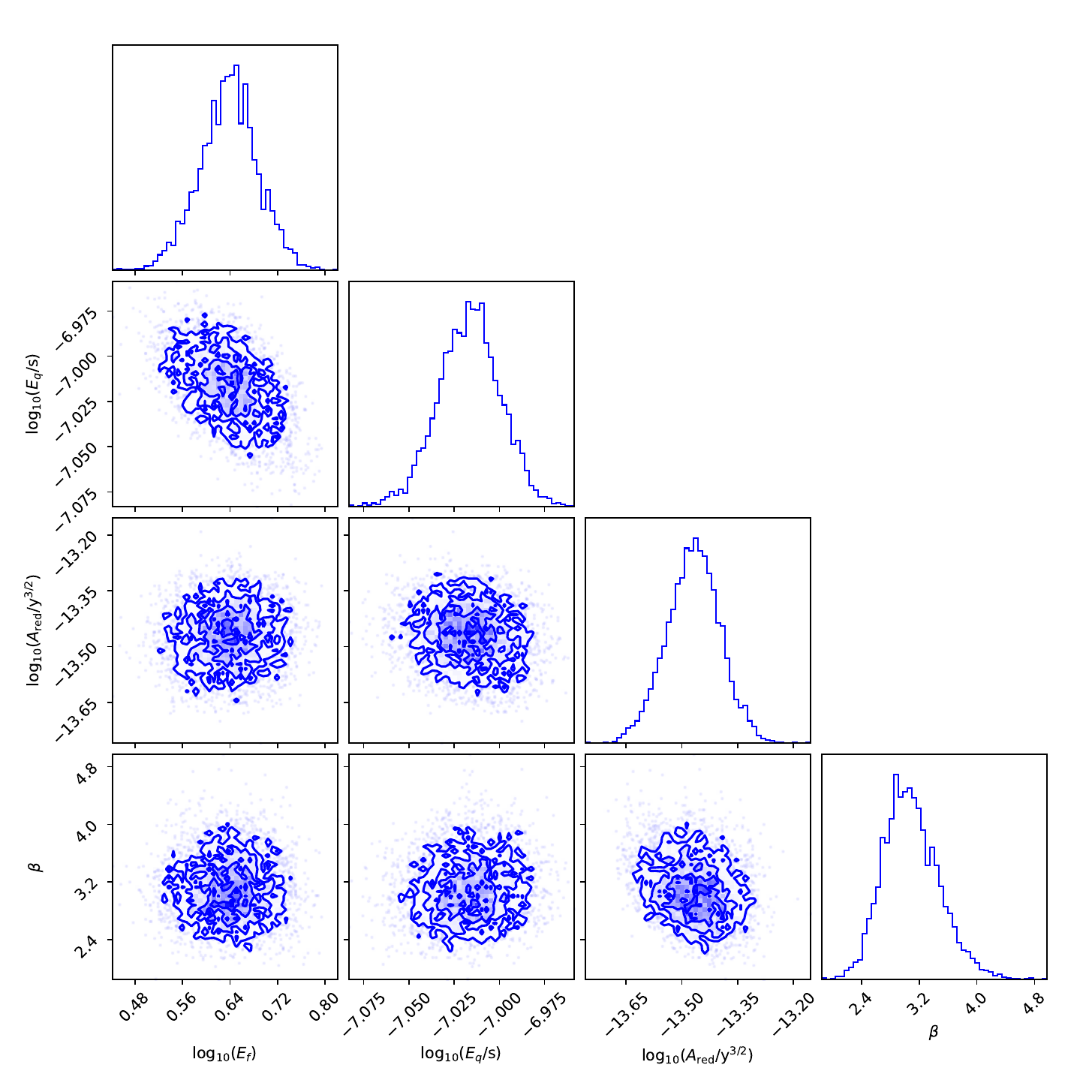}{0.48\textwidth}{(c) IFA-MTM}
          \fig{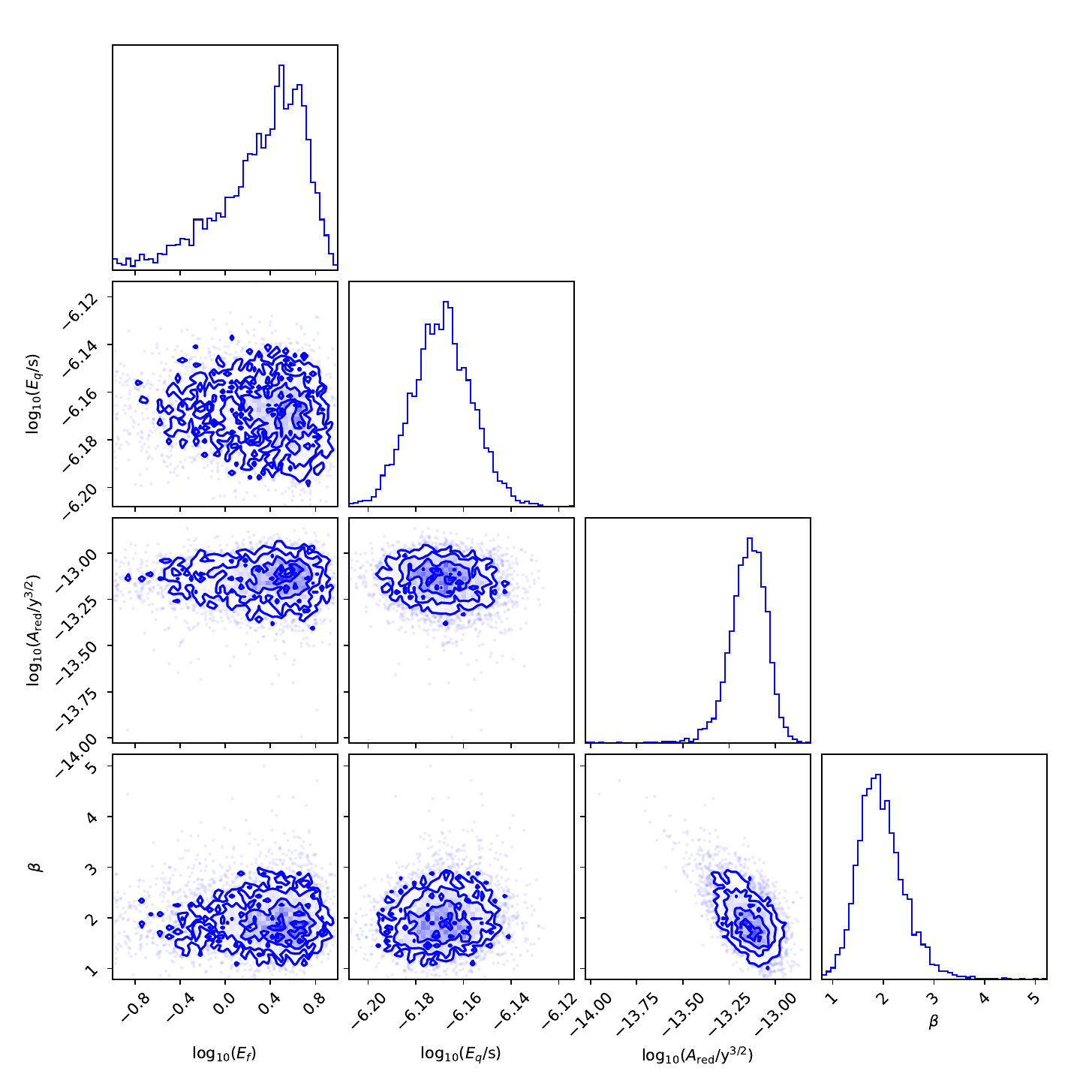}{0.48\textwidth}{(d) IFA-STM}} 
\caption{Noise model parameter distributions for PSR~J0437$-$4715. The two-dimensional posterior surfaces and one-dimensional marginal distributions for red and white noise parameters are derived from the \textsc{MultiNest} chains generated by \textsc{TempoNest} during joint parameter space exploration \citep{Lentati_2014}. 
}
\label{fig:corner_plots}
\end{figure*}

\begin{figure}
    \includegraphics[width=0.5\textwidth]{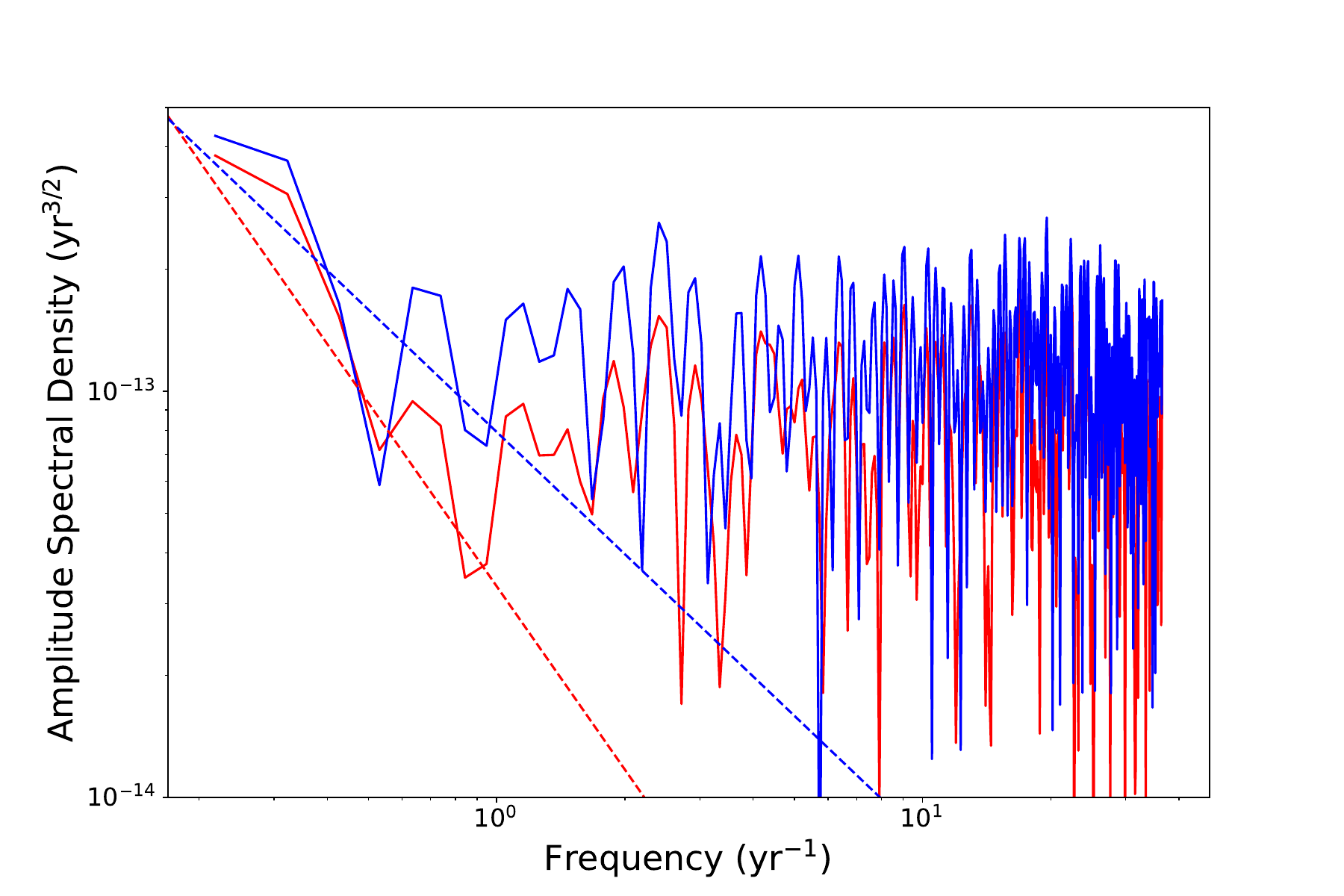}
    \caption{Amplitude spectra of the post-fit residuals of PSR~J0437$-$4715 for the METM-MTM (red) and IFA-STM (blue) data sets.
    The best-fit red noise models output by \textsc{TempoNest} for each data set are indicated by dashed lines with matching colours.}
    \label{fig:amplitude_spectra}
\end{figure}

The IFA-STM and METM-MTM data sets are further compared in Table~\ref{tab:residual_comparison}, which lists \rmswhite\ for each pulsar.
For each pulsar, column 4 lists the {white-noise quotient
\begin{equation}
Q = \frac{\rmswhite\{\mathrm{METM-MTM}\}}{\rmswhite\{\mathrm{IFA-STM}\}}
\label{eqn:rmsquotient}
\end{equation}
that is used to define the percentage improvement, \mbox{$(1-Q)\times100\%$}.
METM-MTM yields significant reductions in \rmswhite;} 
a similar result was found for three pulsars observed with the Nan\c{c}ay Radio Telescope \citep{gcv+23}.
For the MSPs in our sample, the median reduction in white noise is 33\%
and the maximum reduction of 85\% (an impressive factor of 6.6) is observed for PSR~J0437$-$4715.
For this pulsar, we plot the noise model parameter distributions derived from each data set in Figure~\ref{fig:corner_plots} {and compare the amplitude spectra of the residuals in Figure~\ref{fig:amplitude_spectra}}.
%
\begin{table}
\centering
\caption{Comparison of uncertainty-weighted standard deviations of whitened post-fit timing residuals, \rmswhite. Estimates of \rmswhite\ for the IFA-STM and METM-MTM methods are listed in columns 2 and 3, and the {white-noise quotient, $Q$ defined by equation~(\ref{eqn:rmsquotient}) is listed in column 4}.}
\label{tab:residual_comparison}    
\begin{tabular}{c c c c}
\toprule
Pulsar & \multicolumn{2}{c}{\rmswhite\ (ns)} & $Q$ \\
(JNAME) & IFA-STM &  METM-MTM \\ 
\midrule
PSR~J0437$-$4715  &  663 &  100 & 0.15  \\ 
PSR~J0613$-$0200  & 1117 &  583 & 0.52  \\
PSR~J0711$-$6830  &  972 &  754 & 0.78  \\
PSR~J1017$-$7156  &  312 &  223 & 0.71  \\
PSR~J1022+1001    & 1630 &  839 & 0.51  \\
PSR~J1024$-$0719  & 1081 &  654 & 0.60  \\
PSR~J1045$-$4509  & 4702 & 1280 & 0.27  \\
PSR~J1125$-$6014  &  833 &  833 & 1.00  \\
PSR~J1446$-$4701  & 2227 & 1318 & 0.59  \\
PSR~J1545$-$4550  &  785 &  429 & 0.55  \\
PSR~J1600$-$3053  &  341 &  230 & 0.67  \\
PSR~J1603$-$7202  &  747 &  497 & 0.67  \\
PSR~J1643$-$1224  & 1039 &  452 & 0.43  \\ 
PSR~J1713+0747    &  215 &  201 & 0.93  \\ 
PSR~J1730$-$2304  & 1171 &  668 & 0.57  \\
PSR~J1744$-$1134  &  682 &  306 & 0.45  \\
PSR~J1824$-$2452A &  510 &  537 & 1.05  \\
PSR~J1832$-$0836  &  809 &  597 & 0.74  \\
PSR~J1857+0943    &  889 &  573 & 0.64  \\
PSR~J1909$-$3744  &  204 &  200 & 0.98  \\
PSR~J1939+2134    &  104 &  101 & 0.97  \\
PSR~J2124$-$3358  & 2886 & 1320 & 0.46  \\
PSR~J2129$-$5721  & 1194 &  914 & 0.77  \\
PSR~J2145$-$0750  &  684 &  513 & 0.75  \\ 
PSR~J2241$-$5236  &  229 &  214 & 0.93  \\
\bottomrule
\end{tabular}
\end{table}
%
%
%
%

\begin{figure*}
\centering
\includegraphics[width=\textwidth]{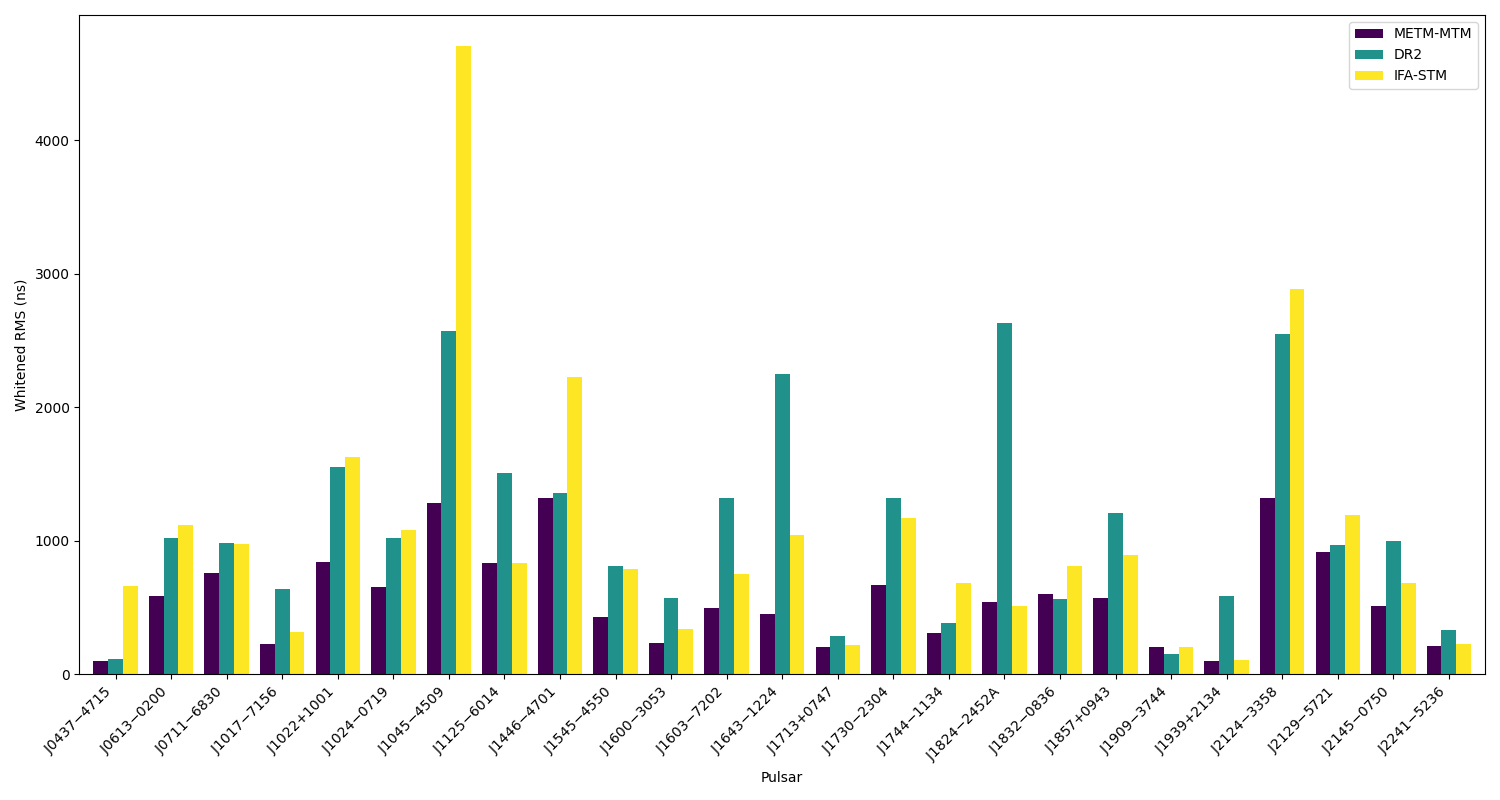}
\caption{Comparison of DR2, METM-MTM, and IFA-STM \rmswhite\ for all pulsars. Each bar represents the \rmswhite\ value obtained using METM-MTM (blue), DR2 (green), and IFA-STM (yellow) for a specific pulsar. Note that DR2 uses the invariant interval \citep{Britton_2000} for PSR~J0437$-$4715.}
\label{fig:bar_graph}
\end{figure*}

Finally, Figure~\ref{fig:bar_graph} depicts the improvement in timing precision for each pulsar by comparing the \rmswhite\ values from PPTA DR2 with those derived in this study using METM-MTM and IFA-STM.
As discussed in more detail in \S\ref{sec:ppta_dr_two}, this visual comparison should be treated as indicative because there are significant differences between the 21-cm data included in the PPTA DR2 analysis and the subset of data analysed in this work.

\section{DISCUSSION}\label{sec:DISCUSSION}
Comparing METM and MTM with conventional methods, we observe a significant reduction in white noise in pulse arrival times across all pulsars in our sample. This is evident from the significantly smaller uncertainty-weighted standard deviations of whitened post-fit timing residuals for METM-MTM compared to IFA-STM (up to 6.6 times smaller; see Table~\ref{tab:residual_comparison} and Table~\ref{tab:timing_results}).
For the majority of PPTA pulsars, the reduction in white noise achieved with METM-MTM exceeds the predicted relative ToA uncertainty between MTM and STM algorithms ($\hat\sigma_{\varphi}$, see Table~\ref{tab:timing_predictions}).
This includes pulsars like PSR~J1744$-$1134 and PSR~J2129$-$5721, 
for which MTM is predicted to perform worse than STM, owing to the multiple correlation between the phase shift and unknown Jones matrix model parameters.
These results indicate that MTM is able to mitigate the impact of polarisation calibration errors, which contribute additional white noise to STM-derived arrival times.

Further supporting this interpretation, the smallest improvements are achieved for pulsars with low susceptibility to calibration error as characterised by $\tau_{\beta}$ (see Table~\ref{tab:timing_predictions}), 
{such as PSR~J1125$-$6014, PSR~J1713$+$0747, PSR~J1909$+$3744, PSR~J1939$-$2134, PSR~J1824$-$2452A, and PSR~J2241$-$5236} (see Table~\ref{tab:residual_comparison}). 
{Similarly, for pulsars that are highly susceptible to calibration error, such as PSR~J0437$-$4715, PSR~J1022$+$1001, PSR~J1045$-$4509, and PSR~J1643$-$1224, METM-MTM significantly reduces \rmswhite.
The largest reduction in \rmswhite, with a quadrature difference of 4.5~$\mu$s between IFA-STM and METM-MTM data sets, is observed for PSR~J1045$-$4509.  This pulsar also has the greatest susceptibility to calibration error $\tau_{\beta}$.}

{The whitened residuals include a contribution from pulsar-intrinsic jitter, which is not separately accounted in our noise analysis.
The white noise induced by jitter is statistically independent of the radiometer noise that dominates the error of each arrival time estimate; therefore, jitter increases $E_q$.  
Jitter noise varies with pulse phase and, when the sub-pulse structures that cause jitter are broader than the pulse phase bins used to resolve the average profile, jitter noise is correlated between phase bins.  That is, jitter causes the noise in each phase bin to no longer be independent and identically distributed, a fundamental assumption on which the STM and MTM algorithms are based.  Therefore, jitter inflates the reduced $\chi^{2}$ of each template-matching fit and causes underestimation of arrival time error, which in turn increases $E_f$.
In summary, pulsar-intrinsic jitter impacts on the time-of-arrival goodness-of-fit, the error scale factor $E_{f}$ and the error added in quadrature $E_q$.  These useful metrics are discussed in the following three sections.}

\subsection{Time-of-Arrival Goodness-of-Fit}
\label{sec:gof}
The template-matching reduced $\chi^{2}$ heat map (see Figure~\ref{fig:heatmap}) presents the median goodness-of-fit for arrival times derived from each pulsar in our study. To increase the dynamic range of the colour scale in this figure, PSR~J0437$-$4715 is excluded because it exhibits the greatest jitter relative to timing uncertainty, leading to exceptionally large values of reduced $\chi^{2}$.

Calibration errors are expected to increase in the reduced $\chi^{2}$ of each ToA estimate. Therefore, MTM is expected to yield lower reduced $\chi^{2}$ values than yielded by STM. This is observed for several of the pulsars in our data set; most notably, PSR~J1022$+$1001, renowned for its temporal profile variations \citep{Kramer_1999, Ramachandran_2003, Hotan_2004}, has one of the two highest reduced $\chi^{2}$ values for IFA-STM. This high value is significantly reduced when using MTM.
{Similarly, for PSR~J1939$+$2134 and PSR~J1713$+$0747, MTM 
yields lower reduced $\chi^{2}$ values than STM. Notable exceptions include PSR~J1824$-$2452A and PSR~J2124$-$3358, for which MTM increases the median reduced $\chi^{2}$}. 
This warrants further explanation, which is currently only speculative.
{These pulsars may} demonstrate greater intrinsic variability in Stokes~Q, U, and V {than in} Stokes~I, possibly due to switching between orthogonally polarized modes.
{The variability might also originate} in the interstellar medium or ionosphere;
{for example, uncorrected Faraday rotation variations
\citep[e.g.][]{ymh+11} could lead to bandwidth depolarization, a transformation that cannot be represented by a Jones matrix and therefore cannot be modeled by MTM. A more quantitative investigation is beyond the scope of this paper, and presents an opportunity for future research.}

\subsection{Error Scale Factor}
When ToA uncertainties accurately reflect the white noise content of arrival time estimates, $E_{f}$ is expected to be close to unity (i.e., $\log_{10}(E_{f})\sim0$). 
Values of $E_{f} > 1$ may suggest that the algorithm for arrival time estimation systematically underestimates ToA uncertainty. Alternatively, it could result from additional noise sources with amplitudes that are proportional to that of the radiometer noise in the integrated pulse profile.

The interpretation of $E_{f}$ aligns with two main results. Firstly, $E_{f}$ appears to indicate that STM underestimates the uncertainty, as evidenced by the median $E_{f}$ around 1.3 for STM compared to around 1.0 for MTM (see Table~\ref{tab:error_corrections}). However, for some pulsars, $E_{f}$ is dominated by jitter; for instance, PSR~J0437$-$4715 has the highest $E_{f}$ value of approximately 4 for each model (For IFA-STM, $E_{f}$ is $\sim 2 \pm 3$ which is poorly constrained and consistent with 4.).  This pulsar displays the lowest METM-MTM whitened noise ($\rmswhite\ \sim 100$~ns) in our data set; it also has the highest level of self-noise relative to the radiometer noise \citep{Oslowski_2011, Parthasarathy_2021}.

Although jitter dominates $E_{f}$ for some pulsars, the error factor remains a valuable diagnostic measure. For example, our initial analysis of $E_{f}$ highlighted errors in both the Phase Gradient Shift (PGS) algorithm (the default STM algorithm when estimating arrival times using \textsc{psrchive} tools like \textsc{pat}) and the \textsc{psrchive} implementation of MTM. After correcting these errors, as described in Appendix~\ref{app:error}, there were no statistically significant differences between the $E_{f}$ values derived from IFA-calibrated and METM-calibrated data for a given pulsar and arrival time estimation algorithm. 

\subsection{Error Added in Quadrature}
Error added in quadrature, $E_q$, characterises any additional sources of unaccounted white noise with amplitudes that are uncorrelated with that of the radiometer noise estimated from the off-pulse baseline. When ToA uncertainties accurately reflect the noise content of arrival time estimates, $E_q \rightarrow 0$ and $\log_{10}(E_q) \rightarrow -\infty$; therefore, larger $E_q$ values indicate greater unaccounted uncertainty in TOA measurements. 

Out of all of the $E_q$ estimates listed in Table~\ref{tab:timing_results}, we compare those derived from
the IFA-STM and METM-MTM data sets and list the {seven} pulsars with statistically significant differences in Table~\ref{tab:significant_equad_differences}.
Most notably, METM-MTM has reduced $E_q$ for PSR~J1045$-$4509 by an order of magnitude. This pulsar also has the greatest reduction in \rmswhite\ and the largest susceptibility to calibration error as characterised by $\tau_{\beta}$. 
It is also interesting to find PSR~J1744$-$1134 in this subset.  For this pulsar, the theoretical uncertainty for MTM-derived arrival times is predicted to be 1.6 times greater than that of STM-derived arrival times; it is one of only two pulsars for which ${\hat\sigma}_{\varphi} > 1$ (see Table~\ref{tab:timing_predictions}).  However,
PSR~J1744$-$1134 also has the second-largest susceptibility to calibration error relative to timing precision ($\tau_\beta / \sigma_\tau \sim 0.28$ in Table~\ref{tab:timing_predictions}).

Across the entire set of pulsars, the median $E_q$ for IFA-STM is 330~ns, which is approximately two times the median $E_q$ of 160~ns for METM-MTM (see Table~\ref{tab:error_corrections}), and there is moderate correlation between $\Delta E_q$ and $\tau_\beta$ (Pearson correlation coefficient $r \sim 0.58$).  
There is also very high correlation between $\Delta E_q$ and the differences in the weighted standard
deviations of the whitened residuals, $\Delta\rmswhite$ ($r \sim 0.92$),
and moderate correlation between $\tau_\beta$ and $\Delta\rmswhite$ ($r \sim 0.63$).
However, it should be noted that the correlations between these variables are reduced when
PSR~J1045$-$4509 is omitted from the data set.  In this case, the correlation between $\Delta E_q$ and $\tau_\beta$ is low ($r \sim 0.37$), the correlation between $\Delta E_q$ and $\Delta\rmswhite$ is moderate ($r \sim 0.60$), and the correlation between $\Delta\rmswhite$ and $\tau_\beta$ is low ($r \sim 0.45$).

As both $E_q$ and \rmswhite\ are significantly reduced by METM-MTM and both are correlated with $\tau_\beta$,
we conclude that polarization calibration errors contribute significantly to the additional white noise in pulsar timing data.  

%
%
%

\subsection{Red Noise}

For most of the PPTA pulsars, the results in Table~\ref{tab:timing_results} show no significant differences in red noise model parameters across the four data sets considered.
The IFA-STM and METM-STM data sets for PSR~J1600$-$3053 and the IFA-STM data set for PSR~J1744$-$1134 appear to have marginally smaller red noise amplitudes and marginally steeper spectra; however, the noise model parameter distribution plots included in Appendix~\ref{app:rejected} (Figure~\ref{fig:corner_plots_1600} and Figure~\ref{fig:corner_plots_1744}) clearly show that, in these three cases, the red noise amplitude and spectral index are both highly covariant and poorly constrained.  Therefore, these results are omitted from further consideration.

In contrast, for PSR~J0437$-$4715, the best-fit red noise model for the IFA-STM data set has a significantly larger amplitude (2.7 $\sigma$ difference) and marginally smaller spectral index (1.7 $\sigma$ difference) than those derived from the other three data sets.  Figure~\ref{fig:corner_plots} verifies that the red noise model parameters are well constrained and only moderately covariant in the IFA-STM and METM-STM data sets.
%
%
%
Both the larger red noise amplitude and smaller spectral index ($\beta=2$) derived from the IFA-STM data set are consistent with the presence of additional $1/f$ noise in the amplitude spectrum of these data.
It is equally likely that additional white noise in the IFA-STM data biases the red noise model parameters.

{Figure~\ref{fig:amplitude_spectra} plots the amplitude spectra of the METM-MTM and IFA-STM post-fit residuals for
PSR~J0437$-$4715 (computed before subtracting the best-fit red noise model).
The amplitude spectral density is the square root of the power spectral density (PSD); therefore,
a PSD characterized by a power law with a spectral index of $\beta$, as defined in equation~\ref{eqn:red_noise_model}, corresponds to a power-law amplitude spectrum with a spectral index of $\beta^\prime = \beta/2$.
The steeper power-law spectrum with slope $\beta^\prime=3/2$ that best fits the METM-MTM data is either buried under additional red noise with slope $\beta^\prime=1$ in the IFA-STM data or buried under additional white noise at key frequencies around yr$^{-1}$ that constrain the slope of the spectrum.

%

It may be possible to differentiate between these two equally 
probable interpretations by analyzing data that span more time and radio frequencies \citep[e.g.][]{Zic_2023}, or by searching for the inter-pulsar correlated systematic timing error induced by polarimetric distortion \citep{Straten_2013}.
This systematic error is a function of the polarized emission from each pulsar,
and is independent of the angular separation between them.
Therefore, instrumental distortion adversely impacts on the sensitivity of a PTA experiment to all moments (monopolar, dipolar, and quadrupolar) in a multipole expansion of residual timing delays.

\subsection{Impact of METM}

In the interest of better understanding which of the two techniques --  METM or MTM -- has the greatest impact on pulsar timing experiments, this section compares four statistical measures (ToA goodness-of-fit, error scale factor, error added in quadrature, and weighted standard deviation of whitened residuals) for the two calibration methods (METM and IFA) and the following section compares these quantities for the two arrival time estimation methods (MTM and STM).

The information in Figure~\ref{fig:heatmap} and its caption can be used to compare the goodness-of-fit of arrival times computed after calibration using either METM or IFA.
First, when comparing IFA-STM with METM-STM (columns 3 and 5), four pulsars stand out as having exceptionally high values of median reduced $\chi^2$ in the IFA-STM data set: PSR~J0437$-$4715, PSR~J1022+1001, PSR~J1713+0747, and PSR~J1939+2134.
For each of these pulsars, METM significantly improves the STM goodness-of-fit.
When comparing IFA-MTM with METM-MTM (columns 2 and 4), a similar reduction in reduced $\chi^2$ is observed only for PSR~J0437$-$4715.  That is, METM generally has little impact on ToA goodness-of-fit when MTM is used.

In Table~\ref{tab:error_corrections}, comparison of the median values of $E_f$ for METM-STM and IFA-STM (bottom two rows) and the median values of $E_f$ for METM-MTM and IFA-MTM (top two rows) shows that METM only marginally reduces the error scale factor for both template matching techniques.  For arrival times derived using STM, the median value of $E_q$ drops by 30\% from 0.33~$\mu$s (IFA) to 0.23~$\mu$s (METM).  For MTM-derived ToAs, there is negligible difference in the median $E_q$ for IFA and METM.  In summary, METM generally has minimal impact on white noise model parameters when MTM is used.


Although METM significantly reduces the \rmswhite\ of STM-derived arrival times for a number of pulsars in Table~\ref{tab:timing_results} (e.g.\ PSR~J0437$-$4715,
PSR~J1022+1001, PSR~J1603$-$3053, and PSR~J1744$-$1134), there are also some pulsars for which METM increases \rmswhite (e.g.\ PSR~J1832$-$0836 and PSR~J2124$-$3358).  Consequently, the median values of \rmswhite\ listed in Table~\ref{tab:error_corrections} show that, with respect to IFA, METM slightly increases the median value of \rmswhite\ by about 30~ns
for arrival times derived using either template matching method.  When comparing \rmswhite\ of MTM-derived arrival times of each pulsar, the differences between METM and IFA are very small.


In summary, METM has negligible impact on TOAs derived using MTM, and only minimal impact on TOAs derived using STM.
METM improves the STM goodness-of-fit and reduces the \rmswhite\ of STM-derived arrival times for a small number of pulsars.
For the STM-derived arrival times of all pulsars, METM reduces the median value of $E_q$ by 100~ns, but also increases
the median value of \rmswhite\ by 30~ns.


\subsection{Impact of MTM}

As done in the previous section, the information in Figure~\ref{fig:heatmap} (and its caption) is used to compare the goodness-of-fit of arrival times computed using either MTM or STM.
We start by comparing IFA-MTM with IFA-STM (columns 2 and 3).  For each of the four previously-identified pulsars with exceptionally high values of IFA-STM median reduced $\chi^2$ (PSR~J0437$-$4715, PSR~J1022+1001, PSR~J1713+0747, and PSR~J1939+2134), MTM significantly improves the goodness-of-fit.
For PSR~J0437$-$4715 and PSR~J1022+1001, the improvement due to MTM is greater than the improvement due to METM.  For PSR~J1939+2134, the opposite is observed.
When comparing METM-MTM with METM-STM (columns 4 and 5), the reduced $\chi^2$ values for MTM and STM are similar (difference less than 0.05\%) for most pulsars except PSR~J0437$-$4715, PSR~J1022+1001, and PSR~J2241$-$5236, for which MTM achieves a better fit than STM; and PSR~J1824$-$2452A, PSR~J1939+2134, and PSR~J2124$-$3358, for which STM achieves a better fit than MTM.
Overall, the differences between MTM and STM ToA goodness-of-fit are greatest in the IFA-calibrated data set.

In Table~\ref{tab:error_corrections}, comparison of the median values of $E_f$ for METM-MTM and METM-STM (rows 1 and 3) and the median values of $E_f$ for IFA-MTM and IFA-STM (rows 2 and 4) shows that MTM significantly reduces the error scale factor for both calibration techniques.  For data calibrated using METM, the median value of $E_q$ drops by 30\% from 0.23~$\mu$s (STM) to 0.16~$\mu$s (MTM).  For IFA-calibrated data, the median value of $E_q$ drops by 55\% from 0.33~$\mu$s (STM) to 0.15~$\mu$s (MTM).
In general, MTM significantly reduces the median values of both white noise model parameters regardless of which calibration technique is used.  MTM also significantly reduces the maximum values of $E_q$ for both calibration methods (from 3.85~$\mu$s to 0.95~$\mu$s for IFA; and from 4.61~$\mu$s to 0.99~$\mu$s for METM).

A similar trend is observed in the weighted standard deviations of the whitened residuals.  For both calibration techniques, MTM reduces the median value of \rmswhite\ by 0.3~$\mu$s and the maximum value of \rmswhite\ by 3~$\mu$s.

In summary, MTM has significant impact on TOAs derived from data using either calibration technique.
For data calibrated using either IFA or METM, MTM improves the ToA goodness-of-fit, reduces the median $E_f$,
and reduces both the median and maximum values of $E_q$ and \rmswhite.

\subsection{Comparison with PPTA DR2}
\label{sec:ppta_dr_two}

The values of \rmswhite\ derived in this work are compared with those reported for PPTA DR2 in Figure~\ref{fig:bar_graph}.
Although this visual comparison may usefully indicate those pulsars for which improved techniques can be expected to have greatest impact, some caution and consideration is required when interpreting this plot.
First, the PPTA DR2 20-cm data span longer periods of time and include observations made using a variety of instrumental backends, each of which may require the inclusion of jumps to model unknown time delays.  In contrast, our analysis focuses on only one backend.
%
%
Second, the PPTA DR2 data were calibrated using Measurement Equation Modeling \citep[MEM;][]{Straten_2004} and therefore the IFA-STM results presented in this work are expected to have greater values of \rmswhite\ than those reported for PPTA DR2.  This is particularly true for PSR~J1045$-$4509 and PSR~J1446$-$4701.
Furthermore, for PPTA DR2, PSR~J0437$-$4715 was timed using the polarimetric invariant profile \citep{Britton_2000}, which greatly mitigates the impact of polarization calibration errors \citep{Straten_2001}.  Consequently, for this pulsar, \rmswhite\ for IFA-STM is significantly greater than that of PPTA DR2.
Keeping in mind the above caveats, Figure~\ref{fig:bar_graph} shows that METM-MTM yields better timing precision than PPTA DR2 for 23 of the 25 pulsars studied. The two exceptions are PSR~J1832$-$0836 and PSR~J1909$-$3744, pulsars that have low susceptibility to calibration error as characterised by $\tau_\beta$. 



\section{Conclusion}\label{sec:conclusion}

Compared to conventional approaches for polarization calibration and arrival time estimation, the combination of METM \citep{Straten_2013} and MTM \citep{Straten_2006} significantly reduces white noise in pulse arrival times. For PSR~J0437$-$4715, METM and MTM reduce the best-fit amplitude of the red noise in the timing residuals, either by mitigating additional red noise, or by increasing the accuracy of red-noise model parameter estimates by reducing the bias on these parameters due to white noise.

In this work, we evaluated the impact of METM and MTM on timing precision using four different quantities: the ToA goodness-of-fit, error scale factor $E_f$, error added in quadrature $E_q$, and the weighted standard deviation of the whitened post-fit residuals \rmswhite.
We found that, with respect to the baseline IFA-STM data set, both METM and MTM improve the ToA goodness-of-fit; however, METM generally has little impact when MTM is used.  A similar pattern is observed in fig.~16 of \citet{gcv+23}.

Whereas METM only marginally reduces the median error scale factor, MTM significantly decreases the median value of $E_f$. In general, this indicates that the STM algorithm used in this work typically under-estimates arrival time uncertainty; however, for PSR~J0437$-$4715, $E_f$ is dominated by jitter for both template-matching techniques.
We also found that both METM and MTM significantly reduce the error added in quadrature; however, the median value of $E_q$ differs very little between METM and IFA when MTM is used.  In contrast, MTM significantly decreases both the median and maximum values of $E_q$ for both METM and IFA calibration methods.
Finally, METM was shown to marginally increase the median value of \rmswhite, whereas MTM significantly decreases both the median and maximum values of \rmswhite. (As expected, changes in $E_q$ and \rmswhite\ are highly correlated.)

In summary, MTM significantly improves arrival time precision, regardless of the calibration technique used. This indicates that, as long as the method of calibration is sufficiently accurate to minimize bandwidth depolarization, MTM is able to model any residual calibration errors and mitigate their impact on arrival time estimates.  Therefore, we recommend that MTM should be used for most pulsars in every pulsar timing array experiment.
{Depending on the instrument, achieving sufficiently well-calibrated observations may necessitate use of either MEM or METM.}

Increased arrival time precision has the potential to enhance our ability to detect errors in solar system ephemerides \citep{Vallisneri_2020} and terrestrial time models \citep{Hobbs_2020}, facilitate new or more accurate measurements of pulsar properties \citep[e.g.][]{Straten_2013}, and increase PTA sensitivity to the stochastic GWB.  {Therefore, MTM should be adopted and utilized for all future IPTA data releases.}


\section*{Acknowledgements}
{We are grateful to Bill Coles for helpful advice and to the anonymous referee for constructive criticism that greatly improved the manuscript.}
The Parkes radio telescope is part of the Australia Telescope National Facility (https://ror.org/05qajvd42), which the Australian Government funds for operation as a National Facility managed by CSIRO.
We acknowledge the Wiradjuri people as the Traditional Owners of the Observatory site.


\bibliography{local}{}

\begin{thebibliography}{}
\expandafter\ifx\csname natexlab\endcsname\relax\def\natexlab#1{#1}\fi
\providecommand{\url}[1]{\href{#1}{#1}}
\providecommand{\dodoi}[1]{doi:~\href{http://doi.org/#1}{\nolinkurl{#1}}}
\providecommand{\doeprint}[1]{\href{http://ascl.net/#1}{\nolinkurl{http://ascl.net/#1}}}
\providecommand{\doarXiv}[1]{\href{https://arxiv.org/abs/#1}{\nolinkurl{https://arxiv.org/abs/#1}}}

\bibitem[{{Agazie} {et~al.}(2023{\natexlab{a}})}]{Agazie_2023c}
{Agazie}, G., {et~al.} 2023{\natexlab{a}}, \apjl, 951, L9,
  \dodoi{10.3847/2041-8213/acda9a}

\bibitem[{{Agazie} {et~al.}(2023{\natexlab{b}})}]{Agazie_2023}
---. 2023{\natexlab{b}}, \apjl, 951, L8, \dodoi{10.3847/2041-8213/acdac6}

\bibitem[{{Agazie} {et~al.}(2023{\natexlab{c}})}]{Agazie_2023b}
---. 2023{\natexlab{c}}, \apjl, \dodoi{10.48550/arXiv.2309.00693}

\bibitem[{{Aggarwal} {et~al.}(2019)}]{Aggarwal_2019}
{Aggarwal}, K., {et~al.} 2019, \apj, 880, 116, \dodoi{10.3847/1538-4357/ab2236}

\bibitem[{{Ajello} {et~al.}(2022)}]{Ajello_2022}
{Ajello}, M., {et~al.} 2022, Science, 376, 521–523,
  \dodoi{10.1126/science.abm3231}

\bibitem[{{Alam} {et~al.}(2021{\natexlab{a}})}]{Alam_2021a}
{Alam}, M.~F., {et~al.} 2021{\natexlab{a}}, \apjs, 252, 5,
  \dodoi{10.3847/1538-4365/abc6a1}

\bibitem[{{Alam} {et~al.}(2021{\natexlab{b}})}]{Alam_2021b}
---. 2021{\natexlab{b}}, \apjs, 252, 4, \dodoi{10.3847/1538-4365/abc6a0}

\bibitem[{{Antoniadis} {et~al.}(2022)}]{Antoniadis_2022}
{Antoniadis}, J., {et~al.} 2022, \mnras, 510, 4873,
  \dodoi{10.1093/mnras/stab3418}

\bibitem[{{Antoniadis} {et~al.}(2023{\natexlab{a}})}]{Antoniadis_2023}
---. 2023{\natexlab{a}}, Astronomy \&; Astrophysics, 678, A50,
  \dodoi{10.1051/0004-6361/202346844}

\bibitem[{{Antoniadis} {et~al.}(2023{\natexlab{b}})}]{Antoniadis_2023b}
---. 2023{\natexlab{b}}, Astronomy \&; Astrophysics, 678, A48,
  \dodoi{10.1051/0004-6361/202346841}

\bibitem[{{Arzoumanian} {et~al.}(2014)}]{Arzoumanian_2014}
{Arzoumanian}, Z., {et~al.} 2014, \apj, 794, 141,
  \dodoi{10.1088/0004-637X/794/2/141}

\bibitem[{{Arzoumanian} {et~al.}(2015)}]{abb+15}
---. 2015, \apj, 813, 65, \dodoi{10.1088/0004-637X/813/1/65}

\bibitem[{{Arzoumanian} {et~al.}(2016)}]{Arzoumanian_2016}
---. 2016, \apj, 821, 13, \dodoi{10.3847/0004-637X/821/1/13}

\bibitem[{{Arzoumanian} {et~al.}(2018)}]{Arzoumanian_2018}
---. 2018, \apjs, 235, 37, \dodoi{10.3847/1538-4365/aab5b0}

\bibitem[{{Arzoumanian} {et~al.}(2020)}]{Arzoumanian_2020}
---. 2020, \apjl, 905, L34, \dodoi{10.3847/2041-8213/abd401}

\bibitem[{{Babak} {et~al.}(2016)}]{Babak_2016}
{Babak}, S., {et~al.} 2016, \mnras, 455, 1665, \dodoi{10.1093/mnras/stv2092}

\bibitem[{{Bailes} {et~al.}(2016)}]{Bailes_2016}
{Bailes}, M., {et~al.} 2016, in MeerKAT Science: On the Pathway to the SKA, 11,
  \dodoi{10.22323/1.277.0011}

\bibitem[{{Blanco-Pillado} {et~al.}(2018){Blanco-Pillado}, {Olum}, \&
  {Siemens}}]{Blanco-Pillado_2018}
{Blanco-Pillado}, J.~J., {Olum}, K.~D., \& {Siemens}, X. 2018, Physics Letters
  B, 778, 392, \dodoi{10.1016/j.physletb.2018.01.050}

\bibitem[{Brent(1973)}]{Brent_1973}
Brent, R.~P. 1973, {Algorithms for Minimization without Derivatives}, 1st edn.
  (Englewood Cliffs, New Jersey: Prentice-Hall)

\bibitem[{{Britton}(2000)}]{Britton_2000}
{Britton}, M.~C. 2000, \apj, 532, 1240, \dodoi{10.1086/308595}

\bibitem[{{Burke-Spolaor} {et~al.}(2019)}]{Burke-Spolaor_2019}
{Burke-Spolaor}, S., {et~al.} 2019, \aapr, 27, 5,
  \dodoi{10.1007/s00159-019-0115-7}

\bibitem[{{Caballero} {et~al.}(2018)}]{Caballero_2018}
{Caballero}, R.~N., {et~al.} 2018, \mnras, 481, 5501,
  \dodoi{10.1093/mnras/sty2632}

\bibitem[{{Caleb} {et~al.}(2019)}]{Caleb_2019}
{Caleb}, M., {et~al.} 2019, \mnras, 487, 1191, \dodoi{10.1093/mnras/stz1352}

\bibitem[{{Caprini} {et~al.}(2010){Caprini}, {Durrer}, \&
  {Siemens}}]{Caprini_2010}
{Caprini}, C., {Durrer}, R., \& {Siemens}, X. 2010, \prd, 82, 063511,
  \dodoi{10.1103/PhysRevD.82.063511}

\bibitem[{{Champion} {et~al.}(2010)}]{Champion_2010}
{Champion}, D.~J., {et~al.} 2010, \apjl, 720, L201,
  \dodoi{10.1088/2041-8205/720/2/L201}

\bibitem[{{Chen} {et~al.}(2021)}]{Chen_2021}
{Chen}, S., {et~al.} 2021, \mnras, 508, 4970, \dodoi{10.1093/mnras/stab2833}

\bibitem[{{Coles} {et~al.}(2011)}]{Coles_2011}
{Coles}, W., {et~al.} 2011, \mnras, 418, 561,
  \dodoi{10.1111/j.1365-2966.2011.19505.x}

\bibitem[{{Cordes} \& {McLaughlin}(2019)}]{Cordes_2019}
{Cordes}, J., \& {McLaughlin}, M.~A. 2019, \baas, 51, 447,
  \dodoi{10.48550/arXiv.1903.08653}

\bibitem[{{Cordes} \& {Downs}(1985)}]{Cordes_1985}
{Cordes}, J.~M., \& {Downs}, G.~S. 1985, \apjs, 59, 343, \dodoi{10.1086/191076}

\bibitem[{{Cordes} \& {Shannon}(2010)}]{Cordes_Shannon_2010}
{Cordes}, J.~M., \& {Shannon}, R.~M. 2010, arXiv e-prints, arXiv:1010.3785,
  \dodoi{10.48550/arXiv.1010.3785}

\bibitem[{{Demorest} {et~al.}(2013)}]{Demorest_2013}
{Demorest}, P.~B., {et~al.} 2013, \apj, 762, 94,
  \dodoi{10.1088/0004-637X/762/2/94}

\bibitem[{{Desvignes} {et~al.}(2016)}]{Desvignes_2016}
{Desvignes}, G., {et~al.} 2016, \mnras, 458, 3341, \dodoi{10.1093/mnras/stw483}

\bibitem[{{Detweiler}(1979)}]{Detweiler_1979}
{Detweiler}, S. 1979, \apj, 234, 1100, \dodoi{10.1086/157593}

\bibitem[{{Dolch} {et~al.}(2021)}]{Dolch_2021}
{Dolch}, T., {et~al.} 2021, \apj, 913, 98, \dodoi{10.3847/1538-4357/abf48b}

\bibitem[{{Edwards} {et~al.}(2006){Edwards}, {Hobbs}, \&
  {Manchester}}]{Edwards_2006}
{Edwards}, R.~T., {Hobbs}, G.~B., \& {Manchester}, R.~N. 2006, \mnras, 372,
  1549, \dodoi{10.1111/j.1365-2966.2006.10870.x}

\bibitem[{Eilers \& Marx(1996)}]{10.1214/ss/1038425655}
Eilers, P. H.~C., \& Marx, B.~D. 1996, Statistical Science, 11, 89 ,
  \dodoi{10.1214/ss/1038425655}

\bibitem[{{Feroz} {et~al.}(2009){Feroz}, {Hobson}, \& {Bridges}}]{Feroz_2009}
{Feroz}, F., {Hobson}, M.~P., \& {Bridges}, M. 2009, \mnras, 398, 1601,
  \dodoi{10.1111/j.1365-2966.2009.14548.x}

\bibitem[{{Goncharov} {et~al.}(2021)}]{Goncharov_2021}
{Goncharov}, B., {et~al.} 2021, \apjl, 917, L19,
  \dodoi{10.3847/2041-8213/ac17f4}

\bibitem[{Grimstad {et~al.}(2015)}]{SPLINTER}
Grimstad, B., {et~al.} 2015, {SPLINTER: a library for multivariate function
  approximation with splines}, \url{http://github.com/bgrimstad/splinter}

\bibitem[{{Grishchuk}(1975)}]{Grishchuk_1975}
{Grishchuk}, L.~P. 1975, Soviet Journal of Experimental and Theoretical
  Physics, 40, 409

\bibitem[{{Guillemot} {et~al.}(2023){Guillemot}, {Cognard}, {van Straten},
  {Theureau}, \& {G{\'e}rard}}]{gcv+23}
{Guillemot}, L., {Cognard}, I., {van Straten}, W., {Theureau}, G., \&
  {G{\'e}rard}, E. 2023, \aap, 678, A79, \dodoi{10.1051/0004-6361/202347018}

\bibitem[{{Hellings} \& {Downs}(1983)}]{Hellings_Downs_1983}
{Hellings}, R.~W., \& {Downs}, G.~S. 1983, in n/a, Vol.~1, General Relativity
  and Gravitation, Volume 1, ed. B.~{Bertotti}, F.~{de Felice}, \&
  A.~{Pascolini}, 963

\bibitem[{{Hemberger} \& {Stinebring}(2008)}]{Hemberger_2008}
{Hemberger}, D.~A., \& {Stinebring}, D.~R. 2008, \apjl, 674, L37,
  \dodoi{10.1086/528985}

\bibitem[{{Hobbs} {et~al.}(2010)}]{Hobbs_2010}
{Hobbs}, G., {et~al.} 2010, Classical and Quantum Gravity, 27, 084013,
  \dodoi{10.1088/0264-9381/27/8/084013}

\bibitem[{{Hobbs} {et~al.}(2012)}]{Hobbs_2012}
---. 2012, \mnras, 427, 2780, \dodoi{10.1111/j.1365-2966.2012.21946.x}

\bibitem[{{Hobbs} {et~al.}(2019)}]{Hobbs_2019}
---. 2019, Research in Astronomy and Astrophysics, 19, 020,
  \dodoi{10.1088/1674-4527/19/2/20}

\bibitem[{{Hobbs} {et~al.}(2020)}]{Hobbs_2020}
---. 2020, \mnras, 491, 5951, \dodoi{10.1093/mnras/stz3071}

\bibitem[{{Hobbs} {et~al.}(2006){Hobbs}, {Edwards}, \&
  {Manchester}}]{Hobbs_2006}
{Hobbs}, G.~B., {Edwards}, R.~T., \& {Manchester}, R.~N. 2006, \mnras, 369,
  655, \dodoi{10.1111/j.1365-2966.2006.10302.x}

\bibitem[{{Hotan} {et~al.}(2004){Hotan}, {Bailes}, \& {Ord}}]{Hotan_2004}
{Hotan}, A.~W., {Bailes}, M., \& {Ord}, S.~M. 2004, \mnras, 355, 941,
  \dodoi{10.1111/j.1365-2966.2004.08376.x}

\bibitem[{{Hotan} {et~al.}(2005){Hotan}, {Bailes}, \& {Ord}}]{Hotan_2005}
---. 2005, \mnras, 362, 1267, \dodoi{10.1111/j.1365-2966.2005.09389.x}

\bibitem[{{Jenet} \& {Anderson}(1998)}]{Jenet_Anderson_1998}
{Jenet}, F.~A., \& {Anderson}, S.~B. 1998, \pasp, 110, 1467,
  \dodoi{10.1086/316273}

\bibitem[{{Jenet} {et~al.}(2005){Jenet}, {Hobbs}, {Lee}, \&
  {Manchester}}]{Jenet_2005}
{Jenet}, F.~A., {Hobbs}, G.~B., {Lee}, K.~J., \& {Manchester}, R.~N. 2005,
  \apjl, 625, L123, \dodoi{10.1086/431220}

\bibitem[{{Jones} {et~al.}(2017)}]{Jones_2017}
{Jones}, M.~L., {et~al.} 2017, \apj, 841, 125, \dodoi{10.3847/1538-4357/aa73df}

\bibitem[{{Joshi} {et~al.}(2018)}]{Joshi_2018}
{Joshi}, B.~C., {et~al.} 2018, Journal of Astrophysics and Astronomy, 39, 51,
  \dodoi{10.1007/s12036-018-9549-y}

\bibitem[{{Keith} {et~al.}(2013)}]{Keith_2013}
{Keith}, M.~J., {et~al.} 2013, \mnras, 429, 2161, \dodoi{10.1093/mnras/sts486}

\bibitem[{{Kerr} {et~al.}(2020)}]{Kerr_2020}
{Kerr}, M., {et~al.} 2020, \pasa, 37, e020, \dodoi{10.1017/pasa.2020.11}

\bibitem[{{Kobakhidze} {et~al.}(2017){Kobakhidze}, {Lagger}, {Manning}, \&
  {Yue}}]{Kobakhidze_2017}
{Kobakhidze}, A., {Lagger}, C., {Manning}, A., \& {Yue}, J. 2017, European
  Physical Journal C, 77, 570, \dodoi{10.1140/epjc/s10052-017-5132-y}

\bibitem[{{Kramer} \& {Champion}(2013)}]{Kramer_Champion_2013}
{Kramer}, M., \& {Champion}, D.~J. 2013, Classical and Quantum Gravity, 30,
  224009, \dodoi{10.1088/0264-9381/30/22/224009}

\bibitem[{{Kramer} {et~al.}(1999)}]{Kramer_1999}
{Kramer}, M., {et~al.} 1999, \apj, 520, 324, \dodoi{10.1086/307449}

\bibitem[{{Lam} {et~al.}(2017)}]{Lam_2017}
{Lam}, M.~T., {et~al.} 2017, \apj, 834, 35, \dodoi{10.3847/1538-4357/834/1/35}

\bibitem[{{Lasky} {et~al.}(2016)}]{Lasky_2016}
{Lasky}, P.~D., {et~al.} 2016, Physical Review X, 6, 011035,
  \dodoi{10.1103/PhysRevX.6.011035}

\bibitem[{{Lazarus} {et~al.}(2016)}]{Lazarus_2016}
{Lazarus}, P., {et~al.} 2016, \mnras, 458, 868, \dodoi{10.1093/mnras/stw189}

\bibitem[{{Lazarus} {et~al.}(2020)}]{Lazarus_2020}
---. 2020, {CoastGuard: Automated timing data reduction pipeline}, Astrophysics
  Source Code Library, record ascl:2003.008.
\newblock \doeprint{2003.008}

\bibitem[{{Lee}(2016)}]{Lee_2016}
{Lee}, K.~J. 2016, in Astronomical Society of the Pacific Conference Series,
  Vol. 502, Frontiers in Radio Astronomy and FAST Early Sciences Symposium
  2015, ed. L.~{Qain} \& D.~{Li}, 19

\bibitem[{{Lentati} {et~al.}(2014)}]{Lentati_2014}
{Lentati}, L., {et~al.} 2014, \mnras, 437, 3004, \dodoi{10.1093/mnras/stt2122}

\bibitem[{{Lentati} {et~al.}(2015)}]{Lentati_2015}
---. 2015, \mnras, 453, 2576, \dodoi{10.1093/mnras/stv1538}

\bibitem[{{Lentati} {et~al.}(2016)}]{Lentati_2016}
---. 2016, \mnras, 458, 2161, \dodoi{10.1093/mnras/stw395}

\bibitem[{{Lower} {et~al.}(2020)}]{Lower_2020}
{Lower}, M.~E., {et~al.} 2020, \mnras, 494, 228, \dodoi{10.1093/mnras/staa615}

\bibitem[{{Lu} \& {Chipman}(1996)}]{lc96}
{Lu}, S., \& {Chipman}, R.~A. 1996, Journal of the Optical Society of America
  A, 13, 1106, \dodoi{10.1364/JOSAA.13.001106}

\bibitem[{{Lyne} {et~al.}(2010){Lyne}, {Hobbs}, {Kramer}, {Stairs}, \&
  {Stappers}}]{Lyne_2010}
{Lyne}, A., {Hobbs}, G., {Kramer}, M., {Stairs}, I., \& {Stappers}, B. 2010,
  Science, 329, 408, \dodoi{10.1126/science.1186683}

\bibitem[{{Manchester} {et~al.}(2013)}]{Manchester_2013}
{Manchester}, R.~N., {et~al.} 2013, \pasa, 30, e017,
  \dodoi{10.1017/pasa.2012.017}

\bibitem[{{McLaughlin}(2013)}]{McLaughlin_2013}
{McLaughlin}, M.~A. 2013, Classical and Quantum Gravity, 30, 224008,
  \dodoi{10.1088/0264-9381/30/22/224008}

\bibitem[{{Melatos} \& {Link}(2014)}]{Melatos_Link_2014}
{Melatos}, A., \& {Link}, B. 2014, \mnras, 437, 21,
  \dodoi{10.1093/mnras/stt1828}

\bibitem[{{Morello} {et~al.}(2019)}]{Morello_2019}
{Morello}, V., {et~al.} 2019, \mnras, 483, 3673, \dodoi{10.1093/mnras/sty3328}

\bibitem[{{Ng}(2018)}]{Ng_2018}
{Ng}, C. 2018, in n/a, Vol. 337, Pulsar Astrophysics the Next Fifty Years, ed.
  P.~{Weltevrede}, B.~B.~P. {Perera}, L.~L. {Preston}, \& S.~{Sanidas},
  179--182, \dodoi{10.1017/S1743921317010638}

\bibitem[{{Ord} {et~al.}(2004){Ord}, {van Straten}, {Hotan}, \&
  {Bailes}}]{Ord_2004}
{Ord}, S.~M., {van Straten}, W., {Hotan}, A.~W., \& {Bailes}, M. 2004, \mnras,
  352, 804, \dodoi{10.1111/j.1365-2966.2004.07963.x}

\bibitem[{{Os{\l}owski} {et~al.}(2011)}]{Oslowski_2011}
{Os{\l}owski}, S., {et~al.} 2011, \mnras, 418, 1258,
  \dodoi{10.1111/j.1365-2966.2011.19578.x}

\bibitem[{{Parthasarathy} {et~al.}(2019)}]{Parthasarathy_2019}
{Parthasarathy}, A., {et~al.} 2019, \mnras, 489, 3810,
  \dodoi{10.1093/mnras/stz2383}

\bibitem[{{Parthasarathy} {et~al.}(2021)}]{Parthasarathy_2021}
---. 2021, \mnras, 502, 407, \dodoi{10.1093/mnras/stab037}

\bibitem[{{Perera} {et~al.}(2018)}]{Perera_2018}
{Perera}, B.~B.~P., {et~al.} 2018, \mnras, 478, 218,
  \dodoi{10.1093/mnras/sty1116}

\bibitem[{{Perera} {et~al.}(2019)}]{Perera_2019}
---. 2019, \mnras, 490, 4666, \dodoi{10.1093/mnras/stz2857}

\bibitem[{{Phinney}(2001)}]{Phinney_2001}
{Phinney}, E.~S. 2001, arXiv e-prints, astro,
  \dodoi{10.48550/arXiv.astro-ph/0108028}

\bibitem[{{Press} {et~al.}(1992){Press}, {Teukolsky}, {Vetterling}, \&
  {Flannery}}]{Press_1992}
{Press}, W.~H., {Teukolsky}, S.~A., {Vetterling}, W.~T., \& {Flannery}, B.~P.
  1992, {Numerical recipes in C. The art of scientific computing} (n/a)

\bibitem[{{Rajagopal} \& {Romani}(1995)}]{Rajagopal_Romani_1995}
{Rajagopal}, M., \& {Romani}, R.~W. 1995, \apj, 446, 543,
  \dodoi{10.1086/175813}

\bibitem[{{Ramachandran} \& {Kramer}(2003)}]{Ramachandran_2003}
{Ramachandran}, R., \& {Kramer}, M. 2003, \aap, 407, 1085,
  \dodoi{10.1051/0004-6361:20031036}

\bibitem[{{Ransom} {et~al.}(2019)}]{Ransom_2019}
{Ransom}, S., {et~al.} 2019, in Bulletin of the American Astronomical Society,
  Vol.~51, 195, \dodoi{10.48550/arXiv.1908.05356}

\bibitem[{{Reardon}(2021)}]{Reardon_2021}
{Reardon}, D.~J. 2021, MeerGuard.
\newblock \url{https://github.com/danielreardon/MeerGuard}

\bibitem[{{Reardon} {et~al.}(2016)}]{Reardon_2016}
{Reardon}, D.~J., {et~al.} 2016, \mnras, 455, 1751,
  \dodoi{10.1093/mnras/stv2395}

\bibitem[{{Reardon} {et~al.}(2023)}]{Reardon_2023}
---. 2023, \apjl, 951, L6, \dodoi{10.3847/2041-8213/acdd02}

\bibitem[{{Sazhin}(1978)}]{Sazhin_1978}
{Sazhin}, M.~V. 1978, Soviet Astronomy, 22, 36

\bibitem[{{Sesana} {et~al.}(2004){Sesana}, {Haardt}, {Madau}, \&
  {Volonteri}}]{Sesana_2004}
{Sesana}, A., {Haardt}, F., {Madau}, P., \& {Volonteri}, M. 2004, \apj, 611,
  623, \dodoi{10.1086/422185}

\bibitem[{{Shannon} \& {Cordes}(2010)}]{Shannon_Cordes_2010}
{Shannon}, R.~M., \& {Cordes}, J.~M. 2010, \apj, 725, 1607,
  \dodoi{10.1088/0004-637X/725/2/1607}

\bibitem[{{Shannon} {et~al.}(2013{\natexlab{a}})}]{Shannon_2013}
{Shannon}, R.~M., {et~al.} 2013{\natexlab{a}}, Science, 342, 334,
  \dodoi{10.1126/science.1238012}

\bibitem[{{Shannon} {et~al.}(2013{\natexlab{b}})}]{Shannon_2013b}
---. 2013{\natexlab{b}}, \apj, 766, 5, \dodoi{10.1088/0004-637X/766/1/5}

\bibitem[{{Shannon} {et~al.}(2014)}]{Shannon_2014}
---. 2014, \mnras, 443, 1463, \dodoi{10.1093/mnras/stu1213}

\bibitem[{{Shannon} {et~al.}(2015)}]{Shannon_2015}
---. 2015, Science, 349, 1522, \dodoi{10.1126/science.aab1910}

\bibitem[{{Siemens} {et~al.}(2007){Siemens}, {Mandic}, \&
  {Creighton}}]{Siemens_2007}
{Siemens}, X., {Mandic}, V., \& {Creighton}, J. 2007, \prl, 98, 111101,
  \dodoi{10.1103/PhysRevLett.98.111101}

\bibitem[{{Susobhanan} {et~al.}(2020){Susobhanan}, {Gopakumar}, {Hobbs}, \&
  {Taylor}}]{Susobhanan_2020}
{Susobhanan}, A., {Gopakumar}, A., {Hobbs}, G., \& {Taylor}, S.~R. 2020, \prd,
  101, 043022, \dodoi{10.1103/PhysRevD.101.043022}

\bibitem[{{Taylor}(1992)}]{Taylor_1992}
{Taylor}, J.~H. 1992, Philosophical Transactions of the Royal Society of London
  Series A, 341, 117, \dodoi{10.1098/rsta.1992.0088}

\bibitem[{{Taylor} {et~al.}(2016)}]{Taylor_2016}
{Taylor}, S.~R., {et~al.} 2016, \apjl, 819, L6,
  \dodoi{10.3847/2041-8205/819/1/L6}

\bibitem[{{Tiburzi} {et~al.}(2016)}]{Tiburzi_2016}
{Tiburzi}, C., {et~al.} 2016, \mnras, 455, 4339, \dodoi{10.1093/mnras/stv2143}

\bibitem[{{Tukey}(1977)}]{Tukey_1977}
{Tukey}, J.~W. 1977, {Exploratory data analysis} (n/a)

\bibitem[{{Vallisneri} \& {van Haasteren}(2017)}]{Vallisneri_2017}
{Vallisneri}, M., \& {van Haasteren}, R. 2017, \mnras, 466, 4954,
  \dodoi{10.1093/mnras/stx069}

\bibitem[{{Vallisneri} {et~al.}(2020)}]{Vallisneri_2020}
{Vallisneri}, M., {et~al.} 2020, \apj, 893, 112,
  \dodoi{10.3847/1538-4357/ab7b67}

\bibitem[{{van Haasteren} \& {Levin}(2013)}]{Haasteren_Levin_2013}
{van Haasteren}, R., \& {Levin}, Y. 2013, \mnras, 428, 1147,
  \dodoi{10.1093/mnras/sts097}

\bibitem[{{van Haasteren} {et~al.}(2011)}]{Haasteren_2011}
{van Haasteren}, R., {et~al.} 2011, \mnras, 414, 3117,
  \dodoi{10.1111/j.1365-2966.2011.18613.x}

\bibitem[{{van Straten}(2004)}]{Straten_2004}
{van Straten}, W. 2004, \apjs, 152, 129, \dodoi{10.1086/383187}

\bibitem[{{van Straten}(2006)}]{Straten_2006}
---. 2006, \apj, 642, 1004, \dodoi{10.1086/501001}

\bibitem[{{van Straten}(2013)}]{Straten_2013}
---. 2013, \apjs, 204, 13, \dodoi{10.1088/0067-0049/204/1/13}

\bibitem[{{van Straten} {et~al.}(2001)}]{Straten_2001}
{van Straten}, W., {et~al.} 2001, \nat, 412, 158, \dodoi{10.1038/35084015}

\bibitem[{{Verbiest} {et~al.}(2016)}]{Verbiest_2016}
{Verbiest}, J.~P.~W., {et~al.} 2016, \mnras, 458, 1267,
  \dodoi{10.1093/mnras/stw347}

\bibitem[{{Wang} {et~al.}(2022)}]{wsv+22}
{Wang}, J., {et~al.} 2022, \aap, 658, A181, \dodoi{10.1051/0004-6361/202141121}

\bibitem[{{Xu} {et~al.}(2023)}]{Xu_2023}
{Xu}, H., {et~al.} 2023, Research in Astronomy and Astrophysics, 23, 075024,
  \dodoi{10.1088/1674-4527/acdfa5}

\bibitem[{{Xue} {et~al.}(2021)}]{Xue_2021}
{Xue}, X., {et~al.} 2021, \prl, 127, 251303,
  \dodoi{10.1103/PhysRevLett.127.251303}

\bibitem[{{Yan} {et~al.}(2011){Yan}, {Manchester}, {Hobbs}, {van Straten},
  {Reynolds}, {Wang}, {Bailes}, {Bhat}, {Burke-Spolaor}, {Champion},
  {Chaudhary}, {Coles}, {Hotan}, {Khoo}, {Oslowski}, {Sarkissian}, \&
  {Yardley}}]{ymh+11}
{Yan}, W.~M., {Manchester}, R.~N., {Hobbs}, G., {et~al.} 2011, \apss, 335, 485,
  \dodoi{10.1007/s10509-011-0756-0}

\bibitem[{{Yardley} {et~al.}(2010)}]{Yardley_2010}
{Yardley}, D.~R.~B., {et~al.} 2010, \mnras, 407, 669,
  \dodoi{10.1111/j.1365-2966.2010.16949.x}

\bibitem[{{Zhu} {et~al.}(2014)}]{Zhu_2014}
{Zhu}, X.~J., {et~al.} 2014, \mnras, 444, 3709, \dodoi{10.1093/mnras/stu1717}

\bibitem[{{Zic} {et~al.}(2023)}]{Zic_2023}
{Zic}, A., {et~al.} 2023, \pasa, 40, e049, \dodoi{10.1017/pasa.2023.36}

\end{thebibliography}
\bibliographystyle{aasjournal}

\appendix
\section{Arrival time uncertainty corrections}
\label{app:error}

\subsection{Scalar Template Matching Correction}

The {\sc{psrchive}} software used for this study includes two different algorithms for estimating the phase shift between a high-S/N template profile and an observed profile by cross-correlation in the Fourier domain \citep[][hereafter T92]{Taylor_1992}.
The Phase Gradient Shift (PGS) algorithm uses the Van Wijngaarden–Dekker–Brent method  \citep{Brent_1973} to find the phase shift $\tau$ that minimizes an objective $\chi^2$ merit function, then uses the curvature, $\partial^2\chi^2/\partial\tau^2$, 
to calculate the theoretical uncertainty of $\tau$.
The Fourier domain with Markov Chain Monte Carlo (FDM) method minimizes the same objective merit function using the Levenberg-Marquardt algorithm, then optionally uses Markov Chain Monte-Carlo (MCMC) to sample the distribution of $\tau$ and calculate its uncertainty.
The matrix template matching (MTM) algorithm also uses the curvature of $\chi^2$ to compute arrival time uncertainty; therefore, in this work, we experimentally compare MTM and STM using the PGS implementation.  

Both MTM and PGS are expected to underestimate uncertainty at low $S/N$, as demonstrated for the PGS algorithm through simulations \citep[appendix A of][]{Hotan_2005}, mathematical proof \citep[appendix B of][]{abb+15}, and comparative analysis of experimental data \citep{wsv+22}.
However, the original implementation of PGS also overestimates arrival time uncertainty when the fit between the template and observation is poor.
This unexpected result is most obvious when analysing the  PSR~J0437$-$4715 timing residuals. For arrival times estimated using the original PGS implementation, the best-fit estimate of the $E_f$ noise model parameter increased from $\sim$ 1.8 for IFA-calibrated data to $\sim$ 5.9 for METM-calibrated data.  

The apparent increase in uncertainty is an artefact of the original PGS implementation, which computes the formal error of the phase shift based on the incorrect assumption that the reduced $\chi^2$ is unity. %
This is equivalent to assuming that the noise in the post-fit residual profile is equivalent to the radiometer noise in each harmonic ($\sigma$ in eqs.~[A 6] through~[A 11] of T92).
This assumption breaks down when the observed profile is not a good match to the template, as is the case when the total intensity is significantly distorted by residual calibration errors.

Poor template-matching fits in the IFA-calibrated data are expected to increase $\chi^2$; however, owing to the incorrect definition of $\sigma$ in the original PGS implementation, they also artificially inflate the derived arrival time uncertainty ($\sigma_\tau$, defined by eq.~[A 10] of T92).
In contrast, METM calibration significantly reduces distortions to the total intensity profile, and thereby improves the scalar template-matching fit and reduces the arrival time uncertainties yielded by PGS.  Consequently, a larger value of $E_f$ is required to account for things like pulsar-intrinsic jitter in the METM-calibrated data.  Though smaller, a similar inflation of $E_f$ was also observed for PSR~J1022+1001, which is also highly susceptible to polarization calibration errors \citep{Straten_2013}.

The erroneous assumption that the reduced $\chi^2$ equals unity was made optional in the \textsc{psrchive} software
(on 2023 July 4) and this assumption was disabled before reproducing the results and analysis presented in this paper. 

\subsection{Matrix Template Matching Correction}

During our initial analysis of the best-fit noise model parameters produced by \textsc{TempoNest}, we found that all arrival times estimated using MTM had a median $E_f$ of 0.7, indicating that this algorithm systematically \emph{overestimates} arrival time uncertainty by a factor of approximately $\sqrt{2}$.
To better understand the origin of this erroneous scale factor, we revisited both the derivation and 
the implementation of the equations that define the uncertainty of the best-fit MTM phase shift,
first presented in \S\,3.2 of \citet[][hereafter S06]{Straten_2006}.
Here, the covariance matrix that defines the formal uncertainties of the MTM model parameters is given 
by $\bf{C}={\mbf\alpha}^{-1}$, where $\mbf\alpha$ is the curvature matrix defined by eq.~(14) of S06. 
This relationship between $\mbf\alpha$ and $\bf{C}$ follows eq.~(15.5.15) of Numerical Recipes \citep[hereafter NR]{Press_1992}, 
and the definition of $\mbf\alpha$ is based on eqs.~(15.5.8) and~(15.5.11) of NR. 
(In particular, following the discussion in \S\,15.5 of NR, the
term containing a second derivative has been dropped.)
Compared to eq.~(15.5.11) of NR, eq.~(14) of S06 includes an extra factor of 2; however, this factor of 2 was missing in the {\sc{psrchive}} implementation of the calculation.  Erroneously dividing $\mbf\alpha$ by 2 is equivalent to multiplying $\bf{C}$ by 2 and inflating the MTM arrival time uncertainty by $\sqrt{2}$.

A similar factor of 2 error appeared in the {\sc{psrchive}} calculation of the gradient vector, $\mbf\beta$, 
which is also defined by eq.~(15.5.8) of NR. Therefore, for future reference, a complete derivation of both $\mbf\beta$ and $\mbf\alpha$ is provided here, beginning with the merit function defined by eq.~(10) of S06,
\begin{equation}
\chi^2 = \sum_{m=1}^{N/2} \sum_{k=0}^3 { |S_{m,k} -
\trace[\pauli{k}\,\mbf{\rho}^\prime_m]|^2 \varsigma_k^{-2} }.
\label{eqn:merit_two}
\end{equation}
In this equation, $S_{m,k}$ are the complex-valued Fourier transforms of the average pulse profiles
of the observed Stokes parameters, as a function of pulsar spin harmonic $m$ and Stokes parameter index $k$,
$\mbf{\rho}^\prime_m$ is the model coherency matrix for harmonic $m$, 
$\pauli{k}$ are the Hermitian basis matrices, and $\trace[{\bf A}]$ is the trace of matrix $\bf A$. 
Define the observed coherency matrix at harmonic $m$,
\begin{equation}
\mbf{\rho}_m = \frac{1}{2} \sum_{k=0}^3 S_{m,k} \pauli{k},
\end{equation}
such that $S_{m,k} = \trace[\pauli{k}\,\mbf{\rho}_m]$, 
then use the linearity of the matrix trace and the transitivity of matrix multiplication 
to express the difference between observed and model Stokes parameters for harmonic $m$,
\begin{equation}
D_{m,k} = \trace[\pauli{k}\,(\mbf{\rho}_m-\mbf{\rho}^\prime_m)].
\end{equation}
Assuming that the noise in each Stokes parameter is equal, let $\varsigma_k=\varsigma$. 
The first partial derivative of $\chi^2$ with respect to model parameter $\eta_r$ is then
\begin{equation}
\frac{\partial\chi^2}{\partial\eta_r} = -\frac{2}{\varsigma^2}
\sum_{m=1}^{N/2} \sum_{k=0}^3
\real\left[ D_{m,k}
        \trace\left(\pauli{k}\,\frac{\partial\mbf{\rho}^\prime_m}{\partial\eta_r}
             \right)^*
  \right].
\end{equation}
Because $\trace(\pauli{k}\,\mbf{\rho})^*=\trace(\pauli{k}\,\mbf{\rho}^\dagger)$
and both the trace of a matrix and the real part of a complex number are linear,
%
\begin{equation}
\frac{\partial\chi^2}{\partial\eta_r} = -\frac{2}{\varsigma^2}
\sum_{m=1}^{N/2} 
\real\,\trace\left[ \sum_{k=0}^3 D_{m,k} \pauli{k}\,\frac{\partial\mbf{\rho}^{\prime\dagger}_m}{\partial\eta_r}
             \right].
\end{equation}
Combining the definitions of the coherency matrix and the Stokes parameters yields
\begin{equation}
\mbf{\rho} = \frac{1}{2} \sum_{k=0}^3
 \trace(\pauli{k}\,\mbf{\rho}) \pauli{k};
\end{equation}
therefore,
\begin{equation}
    \sum_{k=0}^3 D_{m,k} \pauli{k} =  2 ( \mbf{\rho}_m - \mbf{\rho}^\prime_m )
\end{equation}
and
\begin{equation}
\beta_r \equiv -\frac{1}{2} \frac{\partial\chi^2}{\partial\eta_r} = \frac{2}{\varsigma^2}
\sum_{m=1}^{N/2} 
\real\,\trace\left[ (\mbf{\rho}_m - \mbf{\rho}^\prime_m) \frac{\partial\mbf{\rho}^{\prime\dagger}_m}{\partial\eta_r} \right].
\end{equation}
Taking the partial derivative of $- \beta_r$ with respect to model parameter $\eta_s$ 
(and dropping the term containing a second derivative) yields
\begin{equation}
\label{eqn:curvature_two}
\alpha_{rs} \equiv \frac{1}{2} 
\frac{\partial^2\chi^2}{\partial\eta_r\partial \eta_s}
= \frac{2}{\varsigma^2} \sum_{m=1}^{N/2}
  \real\,\trace\left[ \frac{\partial\mbf{\rho}^{\prime\dagger}_m}{\partial\eta_r}
                 \frac{\partial\mbf{\rho}^\prime_m}{\partial\eta_s}
          \right].
\end{equation}
which is equivalent to eq.~(14) of S06.
Both $\beta_r$ and $\alpha_{rs}$ include a factor of 2 that was missing in the
{\sc{psrchive}} adaptation of the Levenberg-Marquardt algorithm to complex-valued matrices.
This error was corrected (on 2023 July 9) before reproducing the results and analysis presented in this paper. 

\section{Rejected Red-Noise Models}
\label{app:rejected}

{When comparing the four methods applied to each pulsar in Table~\ref{tab:timing_results}, there are slight differences in the best-fit red noise model parameters that mostly fall within the estimated uncertainties.  Exceptions include three data sets (PSR~J1600$-$3053 IFA-STM and METM-STM, and PSR~J1744$-$1134 IFA-STM) that exhibit marginally smaller red noise amplitudes and steeper spectra; however, the following plots show that these parameters are highly covariant and poorly constrained in these three cases. Consequently, these minor differences are not further considered.}

\begin{figure}
\centering
\gridline{\fig{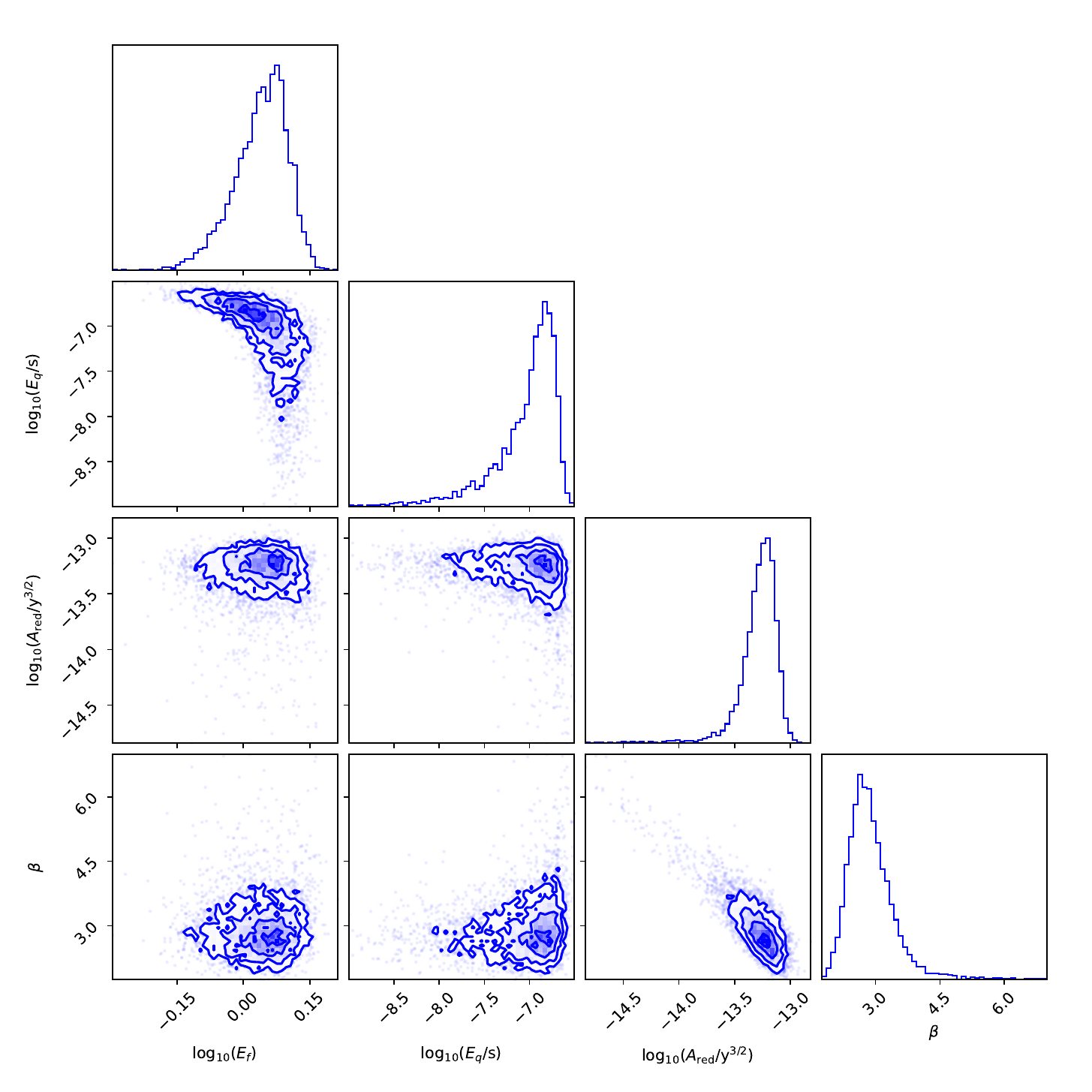}{0.48\textwidth}{(a) METM-MTM}
          \fig{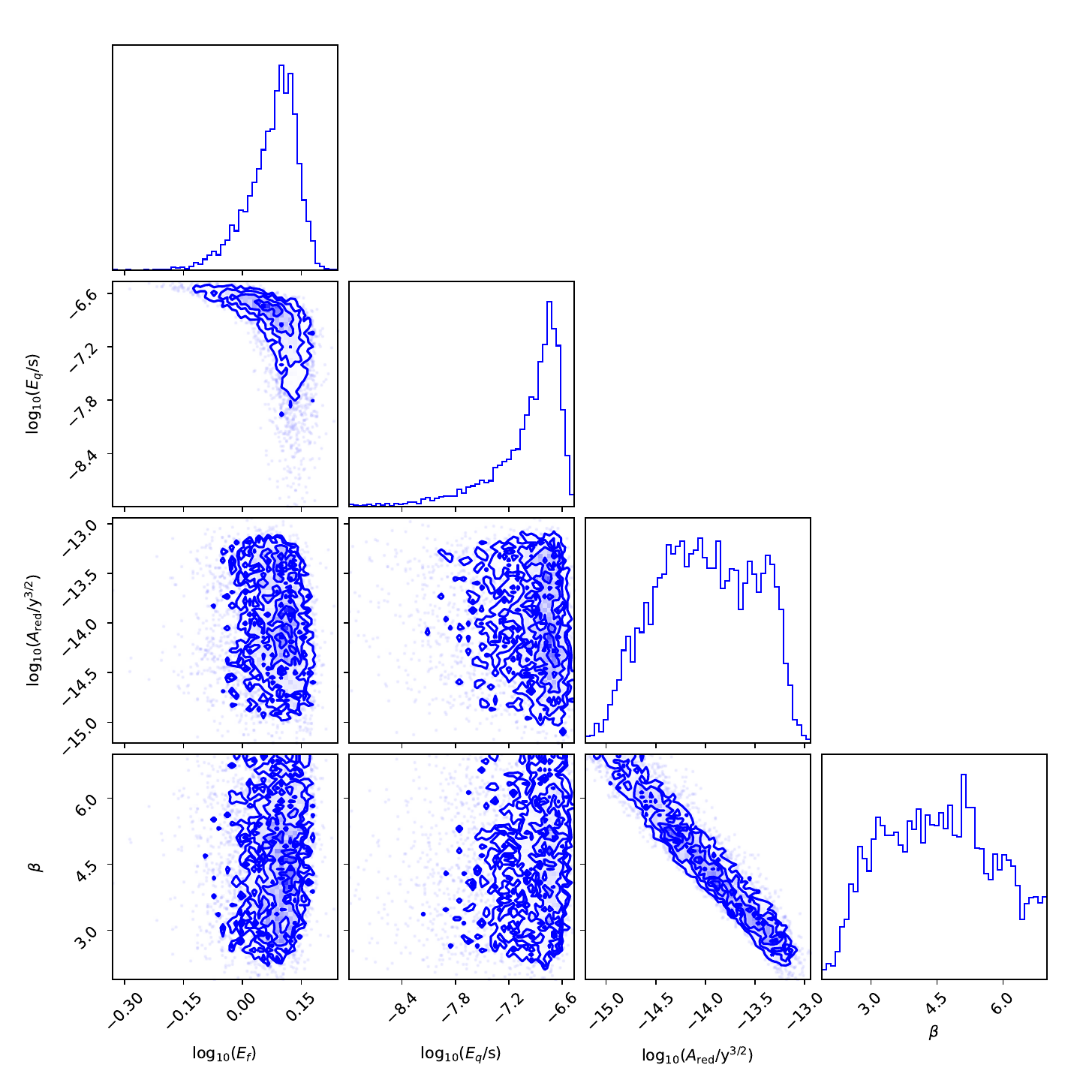}{0.48\textwidth}{(b) METM-STM}}
\gridline{\fig{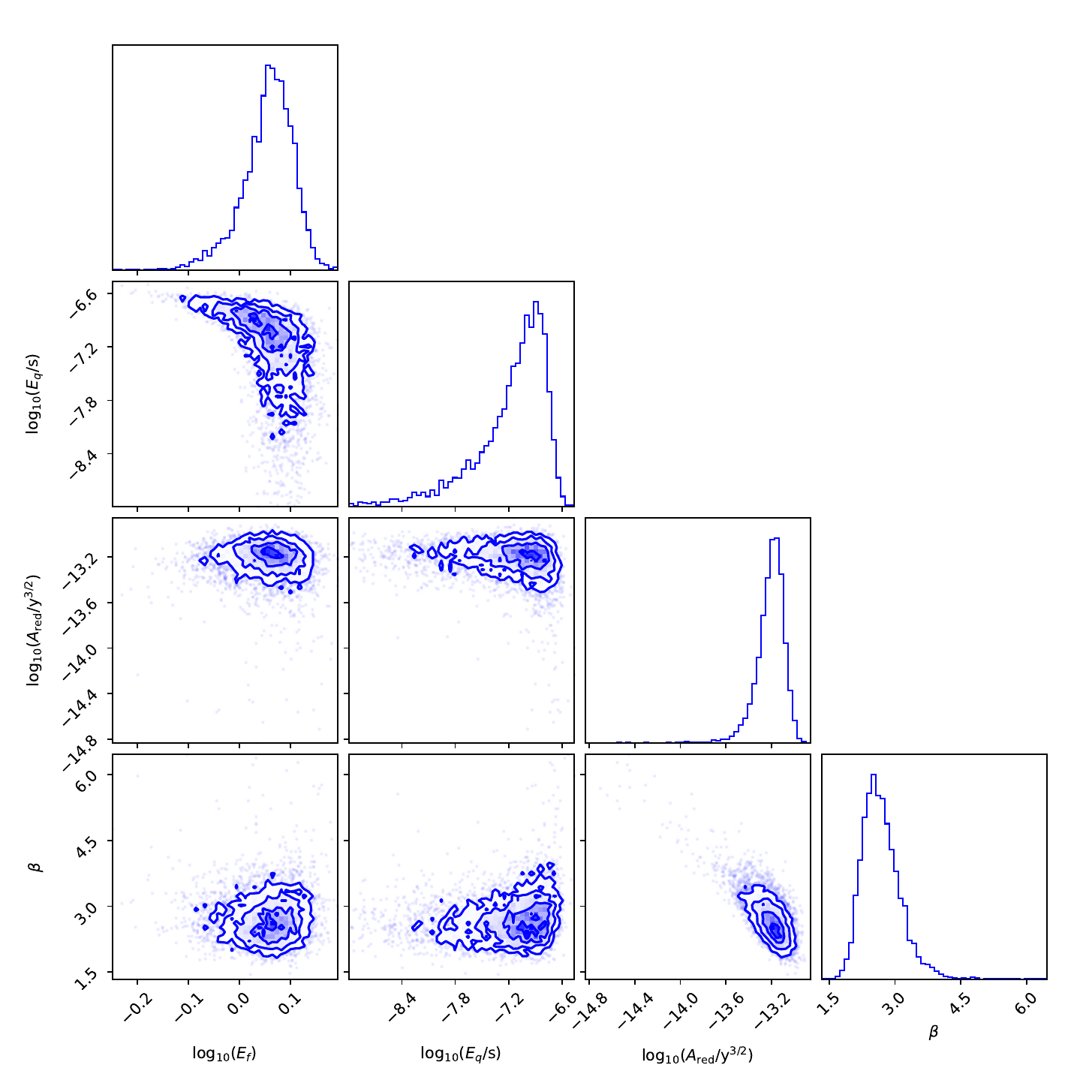}{0.48\textwidth}{(c) IFA-MTM}
          \fig{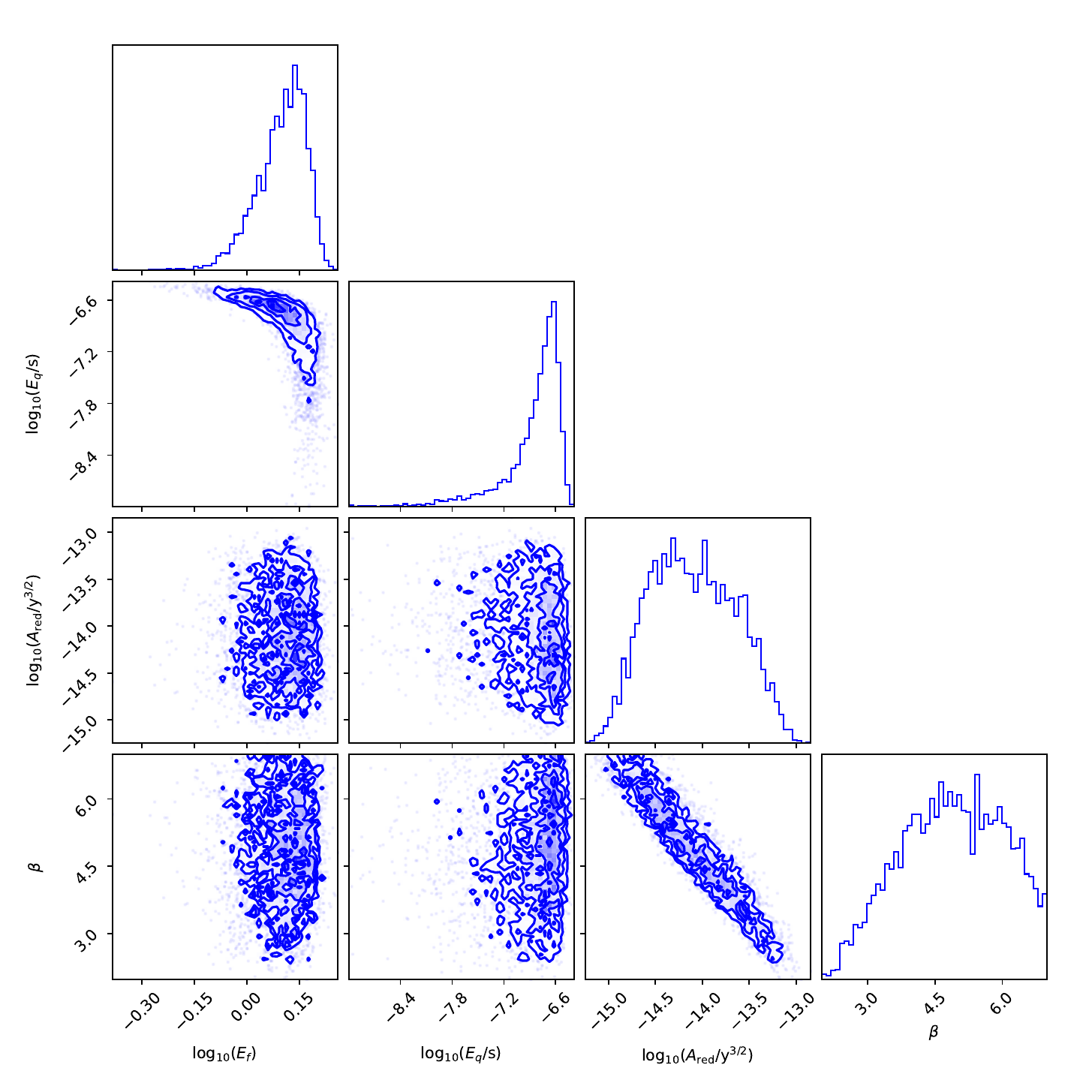}{0.48\textwidth}{(d) IFA-STM}} 
\caption{Noise Model Parameter Distributions for PSR~J1600$-$3053.
See Figure~\ref{fig:corner_plots} for information.}
\label{fig:corner_plots_1600}
\end{figure}

\begin{figure}
\centering
\gridline{\fig{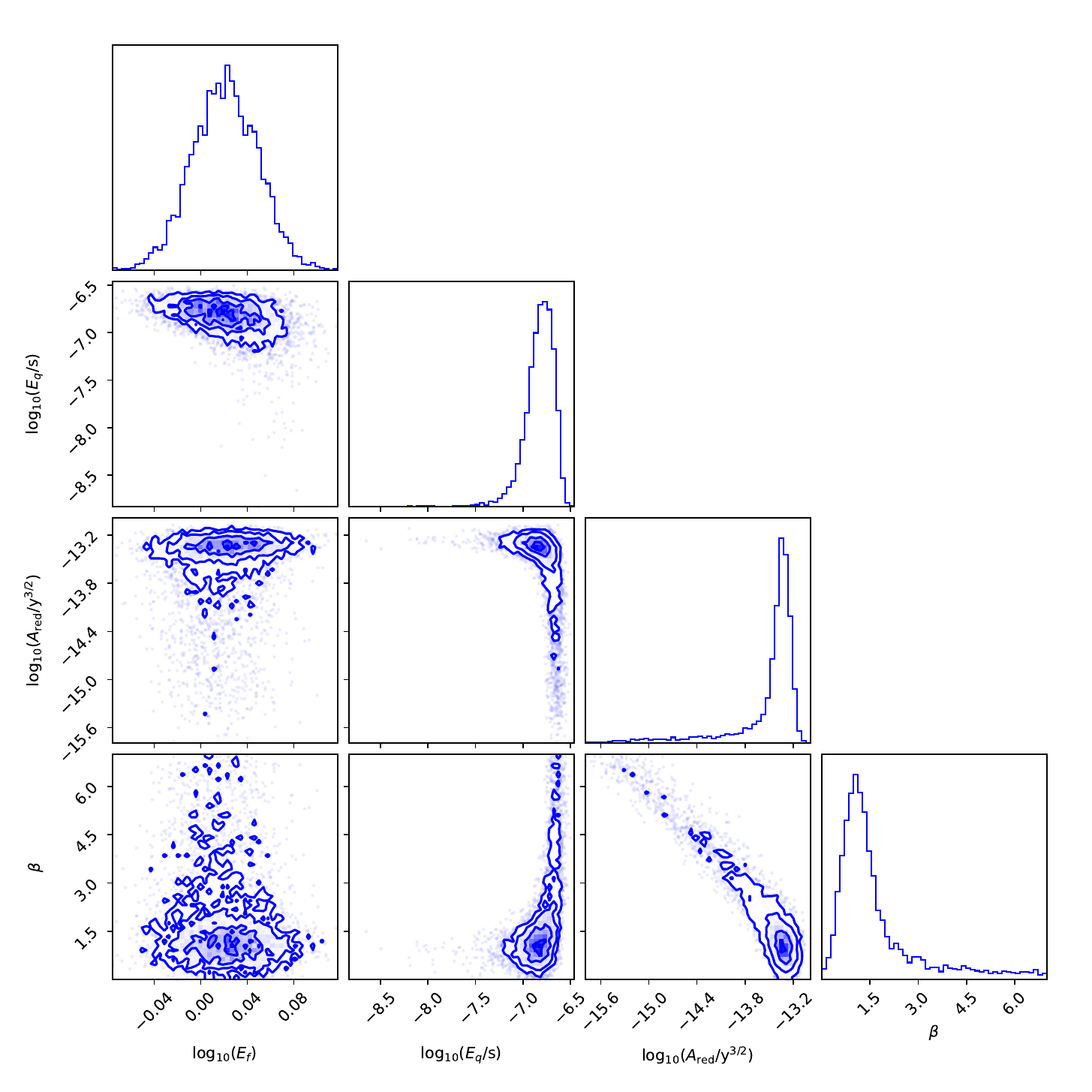}{0.48\textwidth}{(a) METM-MTM}
          \fig{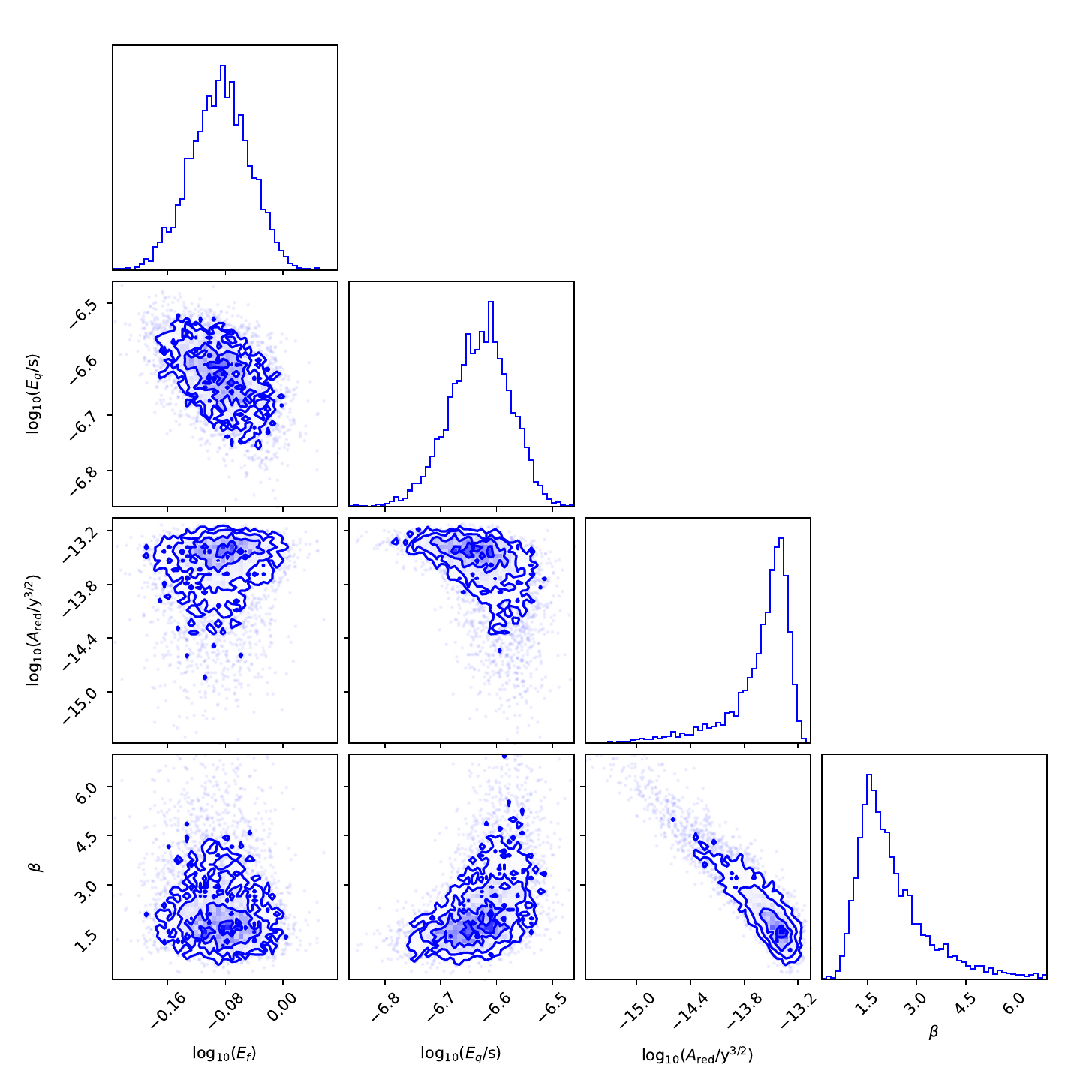}{0.48\textwidth}{(b) METM-STM}}
\gridline{\fig{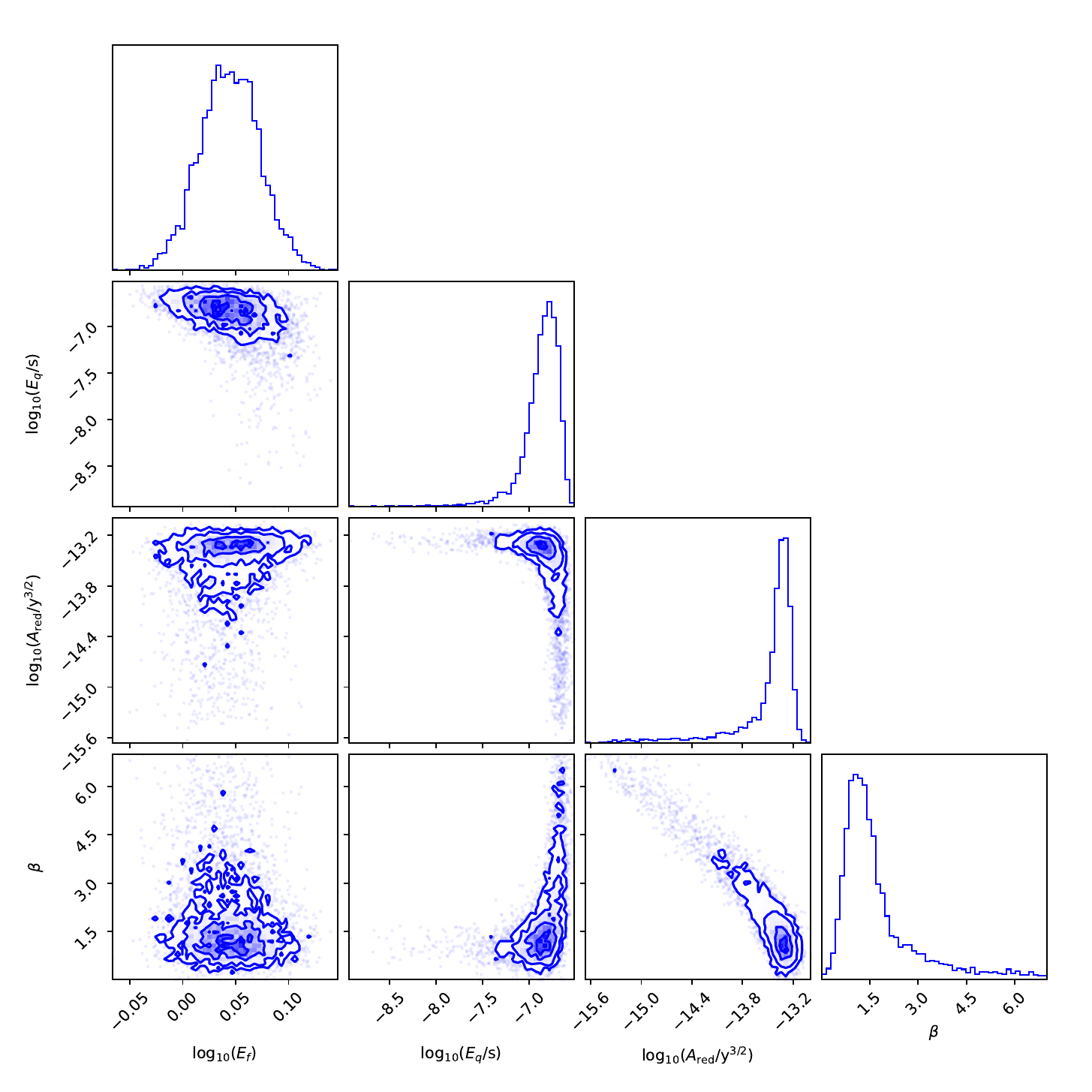}{0.48\textwidth}{(c) IFA-MTM}
          \fig{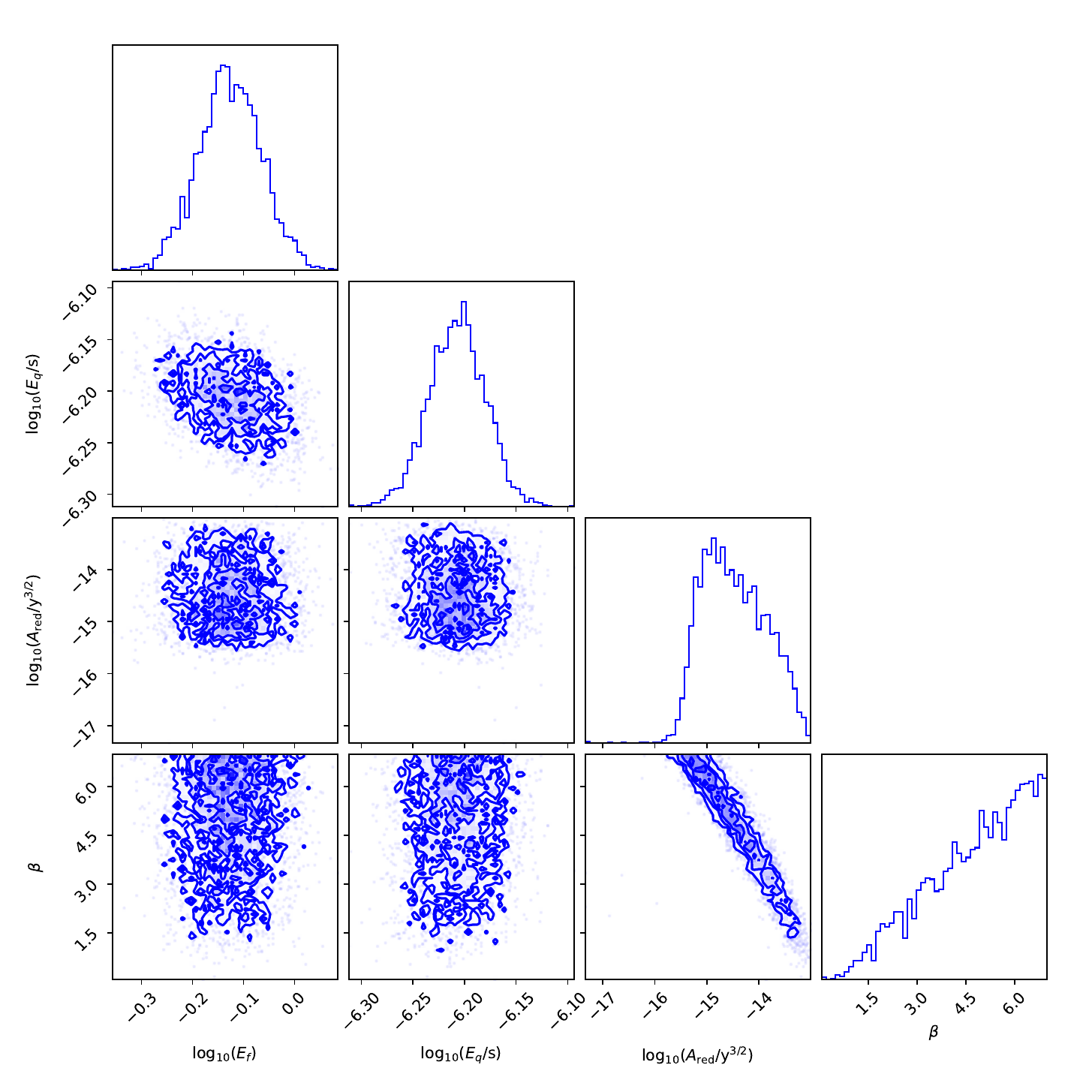}{0.48\textwidth}{(d) IFA-STM}} 
\caption{Noise Model Parameter Distributions for PSR~J1744$-$1134.
See Figure~\ref{fig:corner_plots} for information.}
\label{fig:corner_plots_1744}
\end{figure}


\clearpage
\section{Calibration Steps}
\label{app:steps}

To help others integrate MEM, METM, and MTM into existing pipelines, this appendix provides a 
more detailed description of the commands used for polarimetric calibration.

\subsection{Produce MEM Solutions}

\noindent
{\bf Inputs:}
\begin{enumerate}
    \item Uncalibrated observations of the reference pulsar (e.g., PSR~J0437$-$4715 at Parkes), divided into long sessions, where each session spans at most one day (horizon to horizon) and includes a minimum of 2 hours of observations; and
    \item Uncalibrated observations of the square-wave amplitude-modulated noise diode (CAL) observed prior to each pulsar observation.
\end{enumerate}

\noindent
{\bf Steps:} \\ 
Assuming that the CAL observations are listed in a single flat ASCII file named \texttt{database.txt} (as produced by pac) and that the pulsar observations for a single session are listed in a file named \texttt{session.ls}, the following steps are performed for each session:
\begin{enumerate}
    \item[A.] Calibrate the data using the ideal feed assumption and produce a time-integrated total named \texttt{choose.ar}, which is later passed to pcm for use when choosing the phase bins to include in the MEM fit.
\begin{verbatim}
pac -P -O pac_out -d database.txt -M session.ls
psradd -T -o choose.ar pac_out/*.calibP
\end{verbatim}
    \item[B.] Run pcm in MEM mode.
\begin{verbatim}
pcm $args -c choose.ar -d database.txt -M session.ls
\end{verbatim}
\end{enumerate}

\noindent
where \texttt{\$args} includes the following command-line options: \\

\begin{tabular}{ll}
\texttt{-m bri00e19} &
use Equation (19) of \citealt{Britton_2000} to model the instrumental response \\
\texttt{-k} &
assume that the receptors have equal ellipticities \\
\texttt{-Q} &
model the noise diode as coupled after the orthomode transducer \\
\texttt{-s} &
normalize Stokes parameters by the phase-integrated invariant interval \\
\texttt{-n 64} &
use 64 phase bins as model constraints \\
\texttt{-a 0} &
disable the phase-alignment check \\
\texttt{-K 3.0} &
reject outliers when computing CAL levels \\
\texttt{-step 3.0} &
detect and model steps in instrumental response \\
\texttt{-X 2.0} &
mask channels with $\chi^2 / N_\mathrm{free}$ > 2.0 \\
\texttt{-N} & do not unload calibrated data files
\end{tabular} \\

\noindent
{\bf Outputs:}
\begin{enumerate}
    \item \texttt{pcm.fits} - the best-fit parameters that describe the model of the instrumental response and the polarization of the noise diode (CAL)
    \item \texttt{total.ar} - the calibrated time-integrated full-polarization average pulse profile for this session.
\end{enumerate}
The \texttt{pcm.fits} files should be reviewed, and solutions with any obvious model-fitting problems, errors, or poorly constrained model parameters should be discarded.  

\subsection{Create METM Template Profile}

\noindent
{\bf Inputs:}
\begin{enumerate}
    \item The \texttt{total.ar} files produced for each MEM session.
\end{enumerate}

\noindent
{\bf Steps:}
\begin{enumerate}
    \item[A.] Choose the best MEM solution (criteria described in the paper) and rename its \texttt{total.ar} file to \texttt{chosen.ar}.
    \item[B.] Model and eliminate any temporal variations (e.g. changes in instrumental response, ionospheric Faraday rotation, etc.) before further integrating in time. For each \texttt{total.ar}
\begin{verbatim}
pcm -t8 -n 128 -S chosen.ar -out total.mtm total.ar
\end{verbatim}
    \item[C.] Integrate the calibrated \texttt{total.calib} files output by pcm and the \texttt{chosen.ar} template.
\begin{verbatim}
psradd -T -o pulsar.std chosen.ar */total.calib \
       -j "weight snr fscrunch=1" -phath 0.003
\end{verbatim}
\end{enumerate}

\noindent
{\bf Outputs:}
\begin{enumerate}
    \item \texttt{pulsar.std} - the template to be used for METM
\end{enumerate}

\subsection{Produce METM Solutions}

\noindent
{\bf Inputs:}
\begin{enumerate}
    \item \texttt{pulsar.std} - the high $S/N$, well-calibrated, full-polarization average profile of the reference pulsar
    \item Uncalibrated observations of the reference pulsar (e.g., PSR~J0437$-$4715 at Parkes), divided into short sessions that include a minimum of 1 hour of observations; and
    \item Uncalibrated observations of the square-wave amplitude-modulated noise diode (CAL) observed prior to each pulsar observation.
\end{enumerate}

\noindent
{\bf Steps:} \\ 
Assuming that the CAL observations are listed in a single flat ASCII file named \texttt{database.txt} (as produced by pac) and that the pulsar observations for a single short session are listed in a file named \texttt{session.ls}, the following step is performed for each session:
\begin{verbatim}
     pcm -S pulsar.std -d database.txt -M session.ls $args
\end{verbatim}
\noindent
where \texttt{\$args} includes the following command-line options: \\

\begin{tabular}{ll}
\texttt{-m bri00e19} &
use Equation (19) of \citealt{Britton_2000} to model the instrumental response \\
\texttt{-Q} &
model the noise diode as coupled after the orthomode transducer \\
\texttt{-s} &
normalize Stokes parameters by the phase-integrated invariant interval \\
\texttt{-n 200} &
use 200 harmonics as model constraints \\
\texttt{-K 3.0} &
reject outliers when computing CAL levels \\
\texttt{-step 3.0} &
detect and model steps in instrumental response \\
\texttt{-X 2.0} &
mask channels with $\chi^2 / N_\mathrm{free}$ > 2.0 \\
\end{tabular} \\

\noindent
{\bf Outputs:}
\begin{enumerate}
    \item \texttt{pcm.fits} - the best-fit parameters that describe the model of the instrumental response and the polarization of the noise diode (CAL); and
    \item \texttt{total.ar} - the calibrated time-integrated full-polarization average pulse profile for this session.
\end{enumerate}
The \texttt{pcm.fits} files should be reviewed, and solutions with any obvious model-fitting problems, errors, or poorly constrained model parameters should be discarded.

\subsection{Correct Ionospheric Faraday Rotation}

\noindent
{\bf Inputs:}
\begin{enumerate}
    \item All of the solutions contained in \texttt{pcm.fits} files produced using METM 
\end{enumerate}

\noindent
{\bf Steps:}
Estimate the ionospheric rotation measure contribution and subtract it from each session.
\begin{verbatim}
     pcmrm pcm.fits
\end{verbatim}

\noindent
{\bf Outputs:}
\begin{enumerate}
    \item \texttt{pcm.rms} - the \texttt{pcm.fits} solution with $\sigma_\theta$ corrected for ionospheric Faraday rotation
\end{enumerate}

\subsection{Smooth METM solutions}

\noindent
{\bf Inputs:}
\begin{enumerate}
    \item \texttt{file.ls} - the listing of the \texttt{pcm.rmc} files corrected for ionospheric Faraday rotation using \texttt{pcmrm}.
\end{enumerate}

\noindent
{\bf Steps:}
Spline-smooth the METM solutions using Monte-Carlo cross-validation.
\begin{verbatim}
     smint -p1 -cross -cross-m 4 -cross-f 0.5 -cross-iqr 0 -M files.ls
\end{verbatim}
\noindent
Where the following command-line options specify 

\begin{tabular}{ll}
    \texttt{-p1} & the initial guess for the penalized spline smoothing factor   \\
    \texttt{-cross} & find the optimal spline smoothing factor using Monte-Carlo cross-validation \\
    \texttt{-cross-m 4} & the number of cross-validation iterations \\
    \texttt{-cross-f 0.5} & the fraction of data used to validate the best-fit spline on each iteration \\
    \texttt{-cross-iqr 0} & disables outlier excision using the uncertainty-weighted inter-quartile range \\
    \texttt{-M files.ls} & list of \texttt{pcm.fits} files (corrected for ionospheric Faraday rotation)
\end{tabular} \\

\noindent
{\bf Outputs:}
\begin{enumerate}
    \item \texttt{smint.fits} - the spline-smoothed calibrator solution
    \item \texttt{cal$\_$stokes$\_$fit*.eps}
    \item \texttt{pcal$\_$fit$\_$*.eps}
\end{enumerate}
The postscript files plot the spectrum for each of the smoothed model parameters; each page shows the data extracted from one of the input \texttt{pcm.fits} files (black points with error bars) and the spline fit to the data at this epoch (red line).

\subsection{Calibrate the Pulsar Data}

\noindent
{\bf Inputs:}
\begin{enumerate}
    \item \texttt{metm$\_$database.txt} - database of spline-smoothed METM calibrator solutions
    \item \texttt{cal$\_$database.txt} - database of CAL files.
    \item \texttt{fluxcal$\_$database.txt} - database of FLUXCAL files.
    \item \texttt{uncalibrated.ls} - a list of the pulsar observations to be calibrated
\end{enumerate}

\noindent
{\bf Steps:}
Perform METM calibration.
\begin{verbatim}
pac $args -M uncalibrated.ls \
    -d cal_database.txt -d fluxcal_database.txt -d metm_database.txt
\end{verbatim}
where \texttt{\$args} includes the following command-line options:

\begin{tabular}{ll}
\texttt{-K 3.0} & reject outliers when computing calibrator levels \\
\texttt{-g} & frequency-average the data to match the number of channels of the calibrator \\
\texttt{-m b} & use only calibrators observed before the pulsar \\
\texttt{-S} & use the complete Reception model \\
\texttt{-e cmetm} & extension added to output filenames \\  
\end{tabular} \\

\noindent
{\bf Outputs:}
\begin{enumerate}
    \item calibrated pulsar observations with a new extension, \texttt{*.cmetm}
\end{enumerate}

\subsection{Produce Arrival Time Estimates}

\noindent
{\bf Inputs:}
\begin{enumerate}
    \item calibrated pulsar observations listed in a text file named \texttt{calibrated.ls}
    \item a well-calibrated template profile (e.g. the calibrated observation with the highest $S/N$)
\end{enumerate}

\noindent
{\bf Steps:}
\begin{enumerate}
    \item[A.] Time- and Frequency-average the profile data.
\begin{verbatim}
pam -TF -e TF -M calibrated.ls
\end{verbatim}
    \item[B.] Produce the MTM arrival-time estimates in a \texttt{.tim} file.
\begin{verbatim}
pat -Fpcs pulsar.std -f tempo2 -C gof *.TF > mtm.tim
\end{verbatim}
\end{enumerate}
\noindent
{\bf Outputs:}
\begin{enumerate}
    \item \texttt{mtm.tim} - TEMPO2-formatted file of arrival time estimates produced by MTM.
\end{enumerate}
\label{lastpage}

\end{document}